%% file: ms.tex
\documentclass[sigconf, nonacm]{acmart}

\usepackage{listings}
\usepackage{enumitem}
\usepackage{algorithm}
\usepackage{algpseudocode}
\usepackage{subcaption}
\usepackage{adjustbox}
\usepackage{balance}
\usepackage{caption}
\usepackage{float}
\usepackage[compact]{titlesec}

\usepackage[acronym]{glossaries}
\input{acronym.tex}

\newcommand{\subfigurespace}{\hspace{1mm}}
\titlespacing{\section}{0pt}{3pt}{*0}
\titlespacing{\subsection}{0pt}{2pt}{*0}
\titlespacing{\subsubsection}{0pt}{1pt}{4pt}
\author{Kristalys Ruiz-Rohena}
\affiliation{%
	\institution{Department of Electrical and Computer Engineering \\University of Puerto Rico, Mayag\"uez}
}
\email{kristalys.ruiz@upr.edu}

\author{Manuel Rodriguez-Martinez}
\affiliation{%
	\institution{Department of Computer Science and Engineering \\University of Puerto Rico, Mayag\"uez}
}
\email{manuel.rodriguez7@upr.edu}

\begin{document}
%\title{The Software-Defined Execution Engine for Asymmetric Clouds}
\title{ArcaDB: A Container-based Disaggregated Query Engine for Heterogenous Computational Environments}

\begin{abstract}
Modern enterprises rely on data management systems to collect, store, and analyze vast amounts of data  related with their operations. Nowadays, clusters and hardware accelerators (e.g., GPUs, TPUs) have become a necessity  to scale with the data processing demands in many applications related to social media, bioinformatics, surveillance systems, remote sensing, and medical informatics. Given this new scenario,  the architecture of  data analytics engines must evolve to take advantage of these new technological trends. In this paper, we present ArcaDB: a disaggregated query engine that leverages container technology to place operators at compute nodes that fit their performance profile. In ArcaDB, a query plan is dispatched to worker nodes that have different compute characteristics. Each operator is annotated with the preferred type of compute node for execution, and  ArcaDB ensures that the operator gets picked up by the appropriate workers. We have implemented a prototype version of ArcaDB using Java, Python, and Docker containers. We have also completed a preliminary performance study of this prototype, using images and scientific data. This study shows that ArcaDB can speedup query performance by a factor of 3.5x in comparison with a shared-nothing, symmetric arrangement.

\end{abstract}

\maketitle

\input{introduction.tex}
\input{motivation.tex}

\input{architecture.tex}
\input{implementation.tex}
\input{datamodel.tex}
\input{queryprocess.tex}
\input{performance.tex}

\input{relatedwork.tex}
\input{conclusions.tex}

\input{acknowledgment.tex}

\bibliographystyle{ACM-Reference-Format}
\bibliography{biblio,database,references, databaseold}

\end{document}

%% file: acronym.tex
\newacronym{udf}{UDF}{User Defined Function}
\newacronym{orc}{ORC}{Optimized Columnar Format}
\newacronym{olap}{OLAP}{Online Analytical Processing}
\newacronym{udt}{UDT}{User Defined Type}
\newacronym{fifo}{FIFO}{First-In-First-Out}
\newacronym{os}{OS}{Operating System}
\newacronym{iot}{IoT}{Internet of Things}
\newacronym{ecg}{ECG}{Electrocardiogram}
\newacronym{elt}{ELT}{Extract Load Transform}
\newacronym{vm}{VM}{Virtual Machine}

%% file: introduction.tex
\section{Introduction \label{intro}}
Modern enterprises rely on data management systems to collect, store, and analyze vast amounts of data  related with their operations. These data collections are heterogeneous in nature, featuring a mix of row data, columnar data,
% structured data, semi-structured data, and unstructured data. In addition to row-oriented and column-oriented data, these collections include
images, video,  text, and binary files with numeric values (e.g., results from scientific simulations or experiments).  Data analytics engines, and machine learning tools extract value from these data by providing valuable insights and trends that can be used by key stakeholders to make business decisions.  The size and complex nature of these data sets, coupled with the need for sophisticated analysis,  keep pushing data management systems to their limits in terms of performance and
scalability requirements\cite{10.1145/3524284}.

Nowadays, clusters and hardware accelerators (e.g., GPUs, TPUs, FPGAs) have become a {\em necessity}  to scale with the data processing demands in many applications
related to social media, bioinformatics, surveillance systems, remote sensing, and medical informatics. Modern query engines employ multiple worker nodes that act together to answer a
given query.
Increasingly,  data sets have become cloud-resident because many users prefer the convenience, ease of deployment, elastic scale,  and small upfront costs that cloud-resident data management solutions provide. As a result, data and computation are {\em no longer co-located} in the same cluster. Moreover, container technology (e.g., Docker, Kubernetes)  facilitates rapid application development and deployment over clouds, which makes it possible to move computations to nodes close to the data. At the same time, containers also
make it possible to easily deploy complex, user-defined functions (UDFs) and data types (UDTs) that can be used at various stages in the data processing pipelines.

%\begin{figure}[H]
%	\centering
%	\includegraphics[width=0.4\textwidth]{fig/sharednothing2.png}
%	\caption{Shared-Nothing Architecture}
%	\label{SharedNothing}
%\end{figure}

Given this new scenario,  the architecture of  data analytics engines must evolve to take advantage of these new technological trends.  The standard  shared-nothing architecture \cite{sharednothing},  %shown in Figure \ref{SharedNothing},
with its general purpose nodes running the same software stack everywhere has worked very well on alphanumeric  data (e.g., business data). But this architecture is ill-suited to handle queries that involve UDFs, images, text, or video. Symmetric placement of query operators is no longer advantageous, given the availability of heterogeneous computational resources that provide different price-performance tradeoffs. Previous research has shown the performance advantages obtained by off-loading some complex query operators (e.g. group by, string pattern matching, UDFs )  to multi-CPU systems  and accelerators \cite{You_2013, 10.14778/2732967.2732972, Sitaridi_2015, 10.14778/2994509.2994512, 10.14778/3484224.3484229, 10.14778/3551793.3551809}. But, from an economic point of view, by merely adding GPUs (or other accelerators) to all workers nodes, the user will incur in a high-cost for resources, and not all operators will benefit from them \cite{10.1145/3318464.3380595}.  In such case the shared-nothing architecture is not optimal because it does not take into account the compartmentalization of cloud infrastructure or the possible heterogeneity of compute  resources. 

We argue that, to fully take advantage of these heterogeneous computational resources, we need to completely disaggregate the query engine components \cite{10.1145/3524284}. Each operator in a query plan should be  executed inside the type of compute node(s) that better fits the
performance characteristics  of  said operator. Moreover, operator placement should be {\em asymmetric}: the number and nature of  compute instances that run a given query operator ought to depend on factors such as required response time, budget limits, or reliability requirements. This enables users to deploy their engines on public clouds in a budget-conscious  manner, since there is no need to allocate expensive accelerators (e.g., GPUs) to each worker site.

In this paper, we present ArcaDB: a disaggregated query engine that leverages container technology to place operators at compute nodes that fit the performance profile of each operator. Although engine disaggregation has been explored before \cite{10.1145/1066157.1066201}, our innovation lies in the use of containers as building blocks for workers and accelerators used as query processing units. Unlike previous approaches  \cite{You_2013, 10.14778/2732967.2732972, Sitaridi_2015, 10.14778/3484224.3484229, 10.14778/3551793.3551809} , ArcaDB makes use of multiple nodes and accelerators available in a cloud or cluster environment. In ArcaDB, a query plan is dispatched to a pool of heterogeneous  worker nodes that have different compute characteristics. Each operator is annotated with the preferred type of compute node for execution, and  ArcaDB ensures that the operator gets picked up by the appropriate workers. Thus, a query operator that represents a UDF that computes an image classification task will be assigned to nodes that have GPU. Similarly, in a hash join operation, the operator implementing the probing phase will be  assigned to  high-memory nodes. In contrast, an operator that does a simple alphanumeric selection can be placed on general purpose instances.
With this arrangement, we can maximize the use of special-purpose instances,
%and hardware accelerators,
reduce query response time, and keep the per-query costs under control. As in previous analytics engines (e.g., Spark), ArcaDB focuses on read-only query workloads over cloud resident data. ArcaDB  delegates transactional operations to the underlying 
data lake platforms (e.g., DeltaLake), which are better suited for update management.
We have implemented a prototype version of ArcaDB using Java, Python, Docker containers, and other supporting open-source tools. We have also completed a preliminary performance study of this prototype, using image and  scientific data sets. This study shows that ArcaDB can speedup query performance by a factor of 3.5x in comparison with a shared-nothing, symmetric arrangement. Thus, ArcaDB can help users better meet the performance requirements of their applications.

%The complex nature of these data sets, coupled with the need for sophisticated analysis  keep pushing data management systems for additional requirements in terms of performance and functionality.  As data sets grows into terabyte and petabyte scales, single-server systems struggle to scale and this is when parallel database systems spring into action.  These parallel systems are based on the well-known "shared-nothing architecture", where  a group of server nodes provide both storage and data processing capabilities. Thus, data and query processing are {\em co-located} in the nodes forming the data processing cluster.  Deploying and maintaining these clusters can be a major challenge due to the operational complexity of the database software, the costs for acquiring and maintaining the cluster
%	hardware,  the costs of appropriate physical data center facilities for the cluster, and the costs and difficulty in finding, hiring, and retaining the IT staff necessary to
%	manage the cluster.
%
%Over the past decade, cloud computing has revolutionized the way applications are developed and deployed, particularly those that require a cluster for their
%	operations.

\subsection{Contributions \label{contributions}}
In this paper, we make the following original contributions:
\begin{itemize}[leftmargin=0.4cm]
	 \item  {\bf Disaggregated Container Architecture}: We present the design and prototype implementation of ArcaDB, a data analytics engine that uses containers to disaggregate query executing by placing query plan operators at nodes that match the performance demands of those operators. This arrangement enables better use of accelerators and other special-purpose nodes.
	\item {\bf  Heterogeneous Implementation of Query Operators}: We present the methods by which query operators, UDFs, and UDTs can be implemented in various programming languages/frameworks and be seamlessly used in the same query plan.
	\item {\bf Query Operator Placement}: We introduce  heuristics  to assign operators to specific node types after an initial query plan is developed with a query optimizer.
	\item {\bf Performance Study} : We have implemented a prototype of ArcaDB, and we discuss the performance results obtained on various queries that process images and scientific data.
\end{itemize}
%To the best of our knowledge, no other system provides the same set of  novel features and design points that ArcaDB  has.
%\begin{figure}[h]
%	\centering
%	\includegraphics[width=0.30\textwidth]{fig/blank-space.png}
%\end{figure}

\subsection{Paper Organization \label{organization}}
The rest of this paper is organized as follows. Section \ref{motivation}  provides background and justification for ArcaDB.
%Section \ref{background} provides background information  on data analytics engines to better understand the problem that ArcaDB tackles.
%Section \ref{overview} presents expecific methods used as based ideas to develop the architecture.
In section \ref{architecture} we present an overview of the major components of the system. The implementation steps taken to realize ArcaDB are presented in section \ref{implementation}. The data model and access paths used in ArcaDB are discussed in section \ref{datamodel}. Query processing  is presented in section \ref{query-plan-sec}.
In section \ref{performance} we present our preliminary performance study for  ArcaDB. Related works are presented in section \ref{relworks}.
Finally, our summary of  the paper is shown in section \ref{conclusions}.

%% file: motivation.tex
\section{Motivation \label{motivation}}

\subsection{Example Use Cases \label{usecases}}
The following are example use-cases where ArcaDB could be used to improve query performance.

{\bf Traffic Surveillance}: Consider an app that checks for vehicles that pass red lights in a city. The application receives a stream of  pictures and uses a neural network to detect the car, the license plate, and extract the registration number. Then, it issues a query to an in-memory relational system to fetch the registration information for the car and issue a fine. The application manages several hundred traffic lights and is running  24x7. 
%In this application, there is a record that represents a fine. The record includes the car id, date of violation,  fine amount,
%traffic light, coordinates of the  incident, picture of the car, and picture the license plate. Reports and dashboards are used to inform city officials about the fines %that are issued every month. In this scenario, we see that registration data, which is structured, needs to be integrated with image data and with attributes that are %derived from images. 
While queries that get the registration data for the cars are obtained from an engine that is most likely running on high-memory nodes, the inference operations to detect cars, license plates, and other derived features are better handled by GPU nodes. Thus, this type of application is better served with a platform that distributes query load between accelerators  and  memory optimized instances. 

{\bf Scientific Applications}: The PubChem\footnote{PubChem URL: https://pubchem.ncbi.nlm.nih.gov} 
database is one of the largest databases of chemical molecules and their properties, containing millions of entries.  Chemical compounds can be searched by their  Simplified Molecular-Input Line-Entry System (SMILES) string formula. For example, the  SMILES for 
acetaminophen\footnote{Acetaminophen is a pain reliever medication.} is: 
\begin{verbatim}
     [2H]C1=C(C(=C(C(=C1NC(=O)C)[2H])[2H])O)[2H]
\end{verbatim}
Deep Learning models can be trained 
to extract chemical properties from the SMILES representation since it encodes the molecular structures of the compound.  We could have queries such as:

\begin{lstlisting}[language=SQL] 
	select id,compound_name,enthalpy_energy(smile) 
	from PubChem.Compounds; 
\end{lstlisting} 

In this type of query, {\tt enthalpy\_energy()} is a UDF implemented as a PyTorch model that computes enthalpy, which is a form of energy that can be obtained from the compound. Query performance would be enhanced by offloading these types of operations to hardware accelerators. 

%{\bf IoT}: \acrfull{iot} has emerged as a technology that provides the means to extract, store and analyze data from different \emph{smart devices}.  Big Data systems and \acrfull{elt} data pipelines serve as the processing layer on which \acrshort{iot} systems rely for data management operations. Now, consider an scenario where a device takes an \acrfull{ecg}, sends it to a system that is tracking patients, and after running some analysis, determines if there are any immediate cardiac risks. This information is then transferred to an application that generates alerts to medical experts if any intervention is needed. Depending on the sensitivity of the data and security/privacy requirements, data processing will be limited in terms of where data is stored or moved for analysis. In this case, the data management platform has to be {\em portable}, in the sense that query processing modules need to be dynamically installed at the data location, and use the local processing capabilities to solve queries. Thus, the query functionality moves to the data \cite{DBLP:conf/sigmod/RodriguezR00}. 

\subsection{Background}
\subsubsection{Query Engines for Analytics} 
\begin{figure}[htbp]
	\centering
	\includegraphics[width=0.3\textwidth]{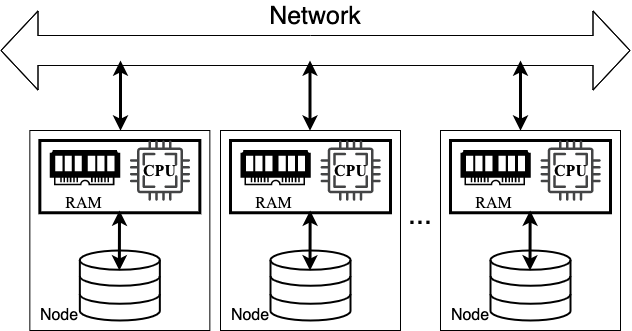}
	\caption{Shared-Nothing Architecture}
	\label{SharedNothing}
\end{figure}

Modern frameworks for data analytics are designed to operate on a cluster of general purpose, off-the-shelf computer nodes.  Examples of these solutions are
Map-Reduce \cite{MapReduce},
Hive \cite{Thusoo:2009:HWS:1687553.1687609}, Spark \cite{10.1145/2723372.2742797}, and Presto \cite{8731547}.
Most of these analytics solutions are based on a shared-nothing architecture \cite{sharednothing}, as shown in Figure \ref{SharedNothing}.  Under this scheme, each computer node receives instructions over the network to perform  individual computational tasks, and performs them using its own private CPU, memory bank, and disk storage.  Typically, an analytics  query engine is composed of a {\em coordinator} node, and a group of {\em worker} nodes (also called ``executors'').   The coordinator serves as the entry point for clients, taking care of tasks such as query parsing, validation, optimization, and plan generation. These activities are guided  by metadata contained in a catalog system. These metadata describe the data sets, integration schemas, available query capabilities, distributions of values  in the data, and many other relevant statistics. 

The coordinator generates an optimized query plan, and ships it to the worker nodes to begin query execution. The plan can either be interpreted, using a Volcano-style iterator interface \cite{344061}, or compiled into a platform specific format \cite{10.14778/2002938.2002940}.  The workers take care of data extraction, schema mapping, query execution, and result delivery. The latter usually happens by placing results into either: 1) an in-memory storage structure, such as a Spark RDDs \cite{10.5555/2228298.2228301} or Alluxio \cite{10.1145/2670979.2670985, Li:EECS-2018-29}, 2) files in a distributed file system (e.g. Hadoop),  or 3) files  in cloud-resident data ``buckets'' (e.g., Amazon S3).  In most systems, every worker node runs the same software stack for data processing, and even the coordinator  uses  modules that are part of this stack. Moreover,  each worker is executing the same query plan, but working on its own shards of the data. We call this {\em symmetric}  operator placement.  Inter-worker collaboration occurs by exchanging data during certain operations such as broadcast joins, or during the shuffling phase in  hash joins.

\subsubsection{Modern Heterogeneous Clouds\label{hetclouds}}
\begin{figure}[htbp]
	\centering
	\includegraphics[width=0.4\textwidth]{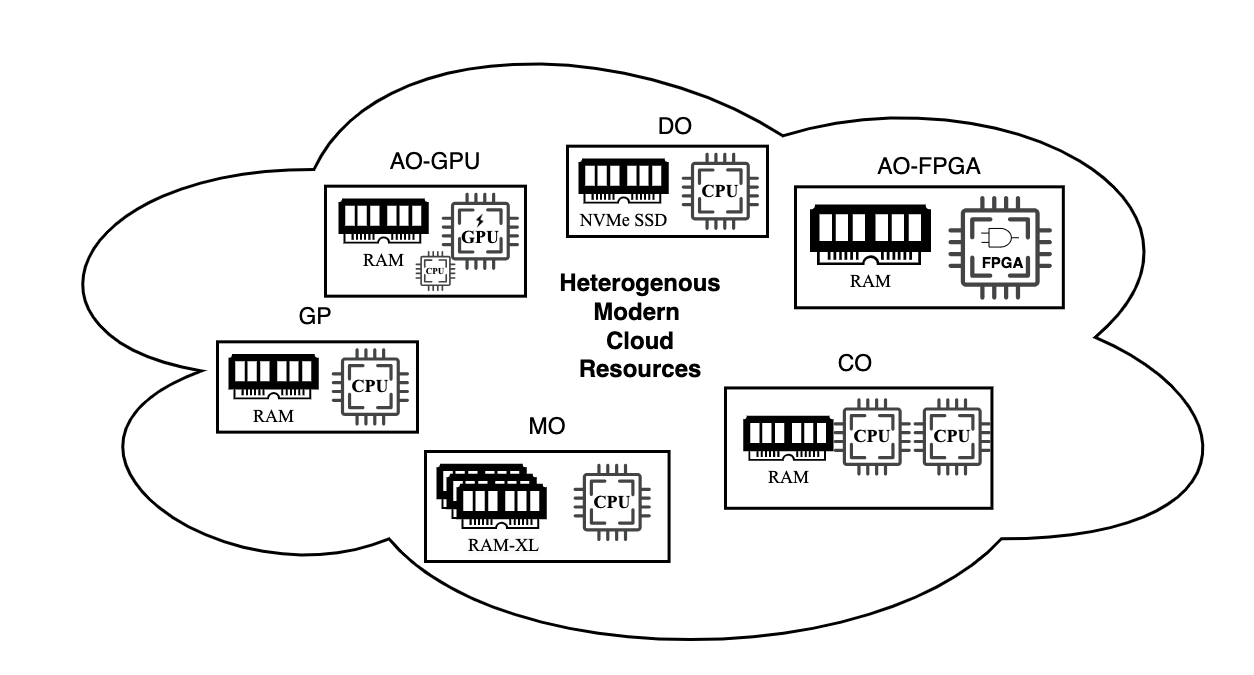}
	\caption{Modern Heterogeneous Clouds}
	\label{modernclouds}
\end{figure}

Modern clouds are {\em asymmetric} in terms of computational capabilities, providing users with a large pool of  diverse node types that are tailored and optimized for specific use-cases. Figure \ref{modernclouds} depicts the most common types of instances, using Amazon AWS and Google Cloud as the basis to derive the illustration. General Purpose (GP) nodes have moderate capabilities in terms of CPU, disk and memory. These instances target the role of  general purpose machines for prototyping and small-scale workloads.  Memory Optimized (MO) nodes feature large and fast banks of main memory, ideally suited for in-memory databases and for high-memory query operators such as hash joins.  Compute Optimized (CO) nodes work best for batch processing workloads, web servers, and compute-bound, user-defined project-select operators \cite{DBLP:conf/sigmod/RodriguezR00}. Accelerator Optimized (AO)  instances feature hardware accelerators to speedup certain operations in comparison with a CPU. AO-FPGA instances contain FPGA  modules that can be programmed to massively accelerate search and pattern-matching operations \cite{10.1145/3035918.3035954} over  strings, or to implement pre-trained machine learning models. AO-GPU instances contain GPU cards to accelerate AI model training, inferences, and to execute aggregate operators with group by clauses \cite{Karnagel2015OptimizingGG}.  Finally, Disk Optimized (DO) nodes contain high-speed solid-state storage (e.g., NVMe)  that provides optimized I/O performance for high-throughput operations such as sorting, sort-merge-join, and to other operations that demand low-latency.

%\begin{figure*}[t]
%	\centering
%	\includegraphics[height=0.19\textheight]{fig/dissagregated-queries.png}
%	\caption{Query Dissagregation}
%	\label{fig:query-dissagregation}
%\end{figure*} 

\subsubsection{Containers and Orchestration}

Heterogeneous resources refer not only to disk and processing capabilities, but  also  to different operating systems, software environments with all dependencies, and API support for applications. As new capabilities emerge, it  becomes a challenge to develop, deploy and maintain applications that run consistently and reliably on these different platforms. In this scenario,  containers provide a solution by bundling an application together with its software dependencies into an enclosed but portable environment that uses the host kernel to execute. 
%The key idea here is that this technology contains all dependencies, configuration files, and functions needed to run a particular application using an underling libraries within  user space. 
The advantages that containers  provide over \acrfull{vm}s are: a) they are generally faster,  and b) are less resource intensive. Containers can also be easily replicated in large quantities and, since they are isolated, various of them can run in the same host by sharing the kernel. For of all these characteristics, containers have become the standard compute organizational unit in modern cloud applications. 

In the context of large scale systems with various components, fleets of containers are harder to manage independently. For such scale, there are containers orchestration tools  like Kubernetes, Docker Swarm, and Openshift. These tools manage many containers across different cloud environments or local servers. They are configured with desired properties like service definition, replication level, resource allocations,  and node type assignment.%, among others. 

\subsection{ArcaDB Overview}
Consider an application that identifies celebrity faces and, based on desired characteristics, retrieves their profile information. We might have the following query that joins pictures of the celebrities with their 
customer profile data in the application.

\begin{lstlisting}[language=SQL,linewidth=\columnwidth]
	select A.id, B.contact_address
	from celeba as A inner join customer as B
	on(A.id=B.id)
	where hasBangs(A.id) and B.id>20  
\end{lstlisting}

This query has a where clause with a predicate that is a UDF  named {\tt hasBangs()}, which is a classifier that identifies if the celebrity in the picture has bangs in his/her hair. The where clause also has a second predicate that discards any record with customer id less or equal to 20. 

\begin{figure*}[htb]
	%	\begin{subfigure}{0.33\textwidth}
		%		\centering
		%		\includegraphics[width=0.7\textwidth]{fig/dissagregatedfig1.png}
		%		\caption{Modern Heterogeneous Clouds}
		%		\label{fig:query-dissagregation1}
		%	\end{subfigure}
	\begin{subfigure}{0.44\textwidth}
		\centering
		\includegraphics[width=0.9\textwidth]{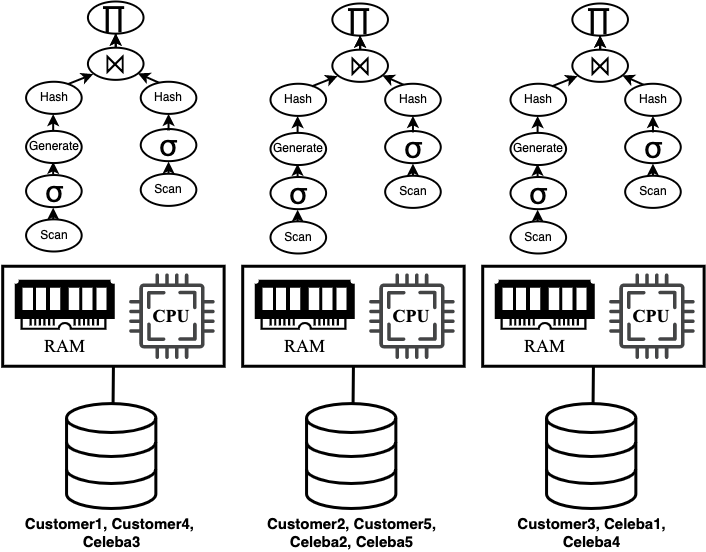}
		\caption{Shared Nothing Query Execution}
		\label{fig:query-dissagregation2}
	\end{subfigure}
	\begin{subfigure}{0.4\textwidth}
		\centering
		\includegraphics[width=0.9\textwidth]{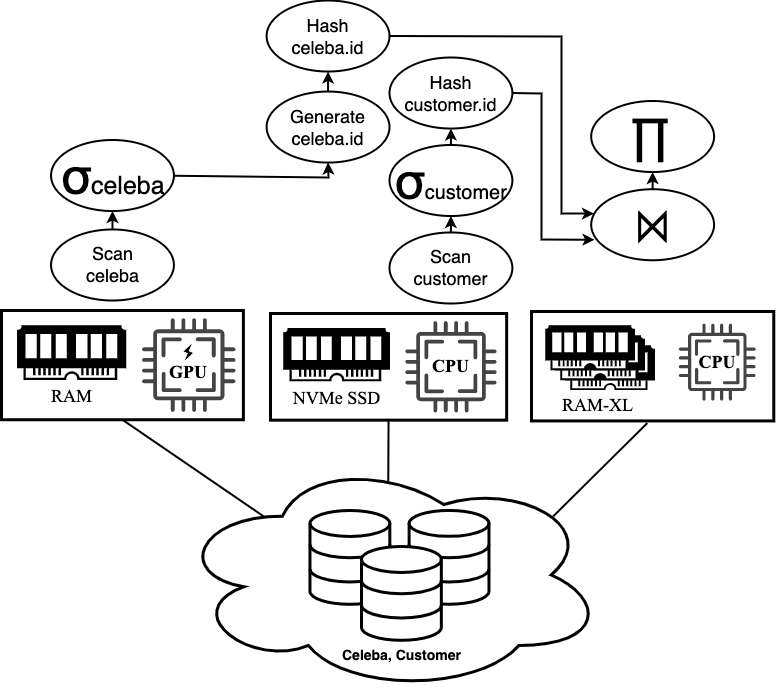}
		\caption{Dissagregated Query Execution}
		\label{fig:query-dissagregation3}
	\end{subfigure}
	\caption{Query Dissagregation}
	\label{fig:three-graphs}
\end{figure*}

\begin{figure}[htbp]
	\centering
	\includegraphics[width=0.18\textwidth]{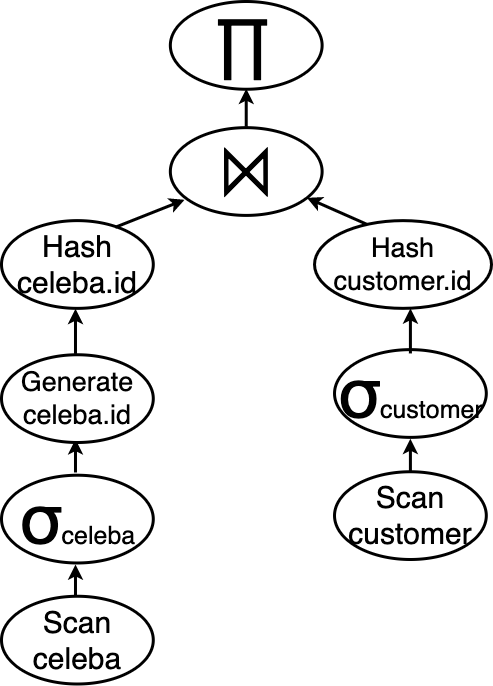}
	\caption{Query Plan for the Celebrities Example. }
	\label{fig:query-dissagregation1}
\end{figure}

For this query, we could generate the query plan that is shown in Figure \ref{fig:query-dissagregation1}.  A hash join method is used to partition and join the data after the predicates on each table in the where clause are evaluated. Then, 
in Figure \ref{fig:query-dissagregation2} we can see how this query plan could be executed in a standard shared-nothing system. Each node receives a copy of the query plan, and runs it on the shards of data it maintains.  Data shuffling will occur as part of the join algorithm. If we were to use accelerators, we would need to have them in each and every node, which increases financial costs to support such queries, thus increasing operational expenses. 
%support for other resources would not be possible.

Figure \ref{fig:query-dissagregation3} shows how this plan is evaluated in ArcaDB. Query operators are placed in nodes that fit their performance profiles. For example,  scanning the pictures and evaluating the {\tt hasBangs()} predicate occurs withing a GPU Node(s). Meanwhile,  scanning the customer table, evaluating the other predicate, and  running the partitioning phase for both tables happens on one or more nodes optimized for I/O. Finally, the probing phase of the join operation occurs 
on one or more  high-memory nodes. 
Intermediate results are placed in an in-memory shared cache that is implemented with Alluxio. With ArcaDB we can still reap the benefits of parallel processing since we can run each operator in multiple nodes. But in addition, we can use accelerators in a budget conscious manner, and  still obtain significant performance gains.

%% file: architecture.tex
\section{ArcaDB  Architecture\label{architecture}}

This section details the main components of ArcaDB  and how the various loosely coupled technologies work together.
%\begin{figure}[h]
%	\centering
%	\includegraphics[width=0.4\textwidth]{fig/GeneralArchi.png}
%	\caption{Conceptual ArcaDB Architecture}
%	\label{GenArcaArquitecture}
%\end{figure}

\subsection{Overview}
ArcaDB features one or more coordination units that manage all aspects of query planning and  processing. Query operators are executed by workers that run on different nodes in the underlying infrastructure. The workers are heterogeneous in the sense that they are specialized to run a specific type of operator on a specific type of node. For example, we can have workers that focus on running \emph{group by} operators or \emph{generalized projections} on GPU nodes.  Likewise, there are workers devoted to running selections or schema mapping operators on the raw data. The number of workers is a configurable parameter  that must be tuned to the specific workloads that will be served with ArcaDB.

\begin{figure}[htb]
	\centering
	\includegraphics[width=0.42\textwidth]{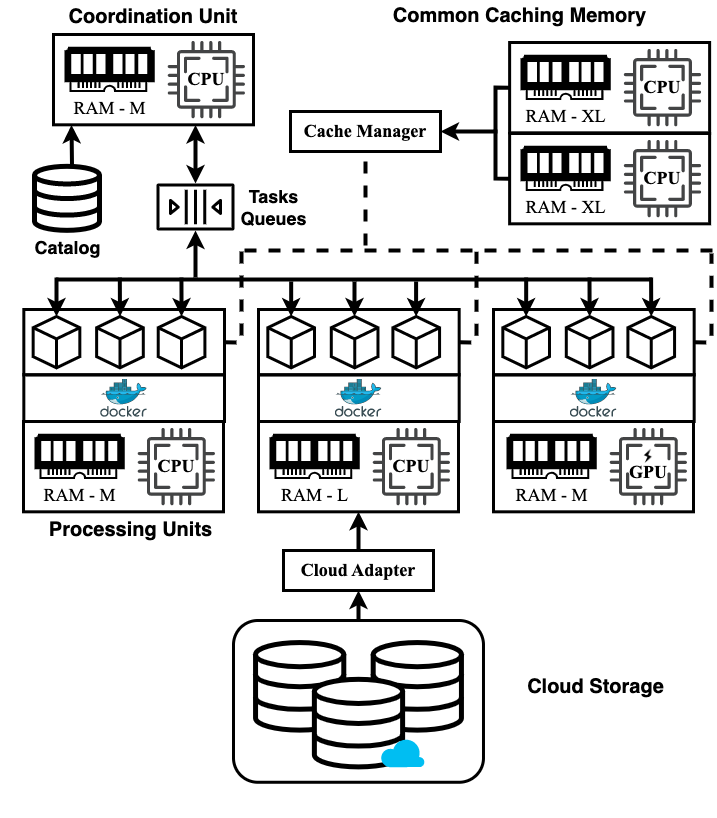}
	\caption{ArcaDB Architecture}
	\label{ArcaArquitecture}
\end{figure}

Implementing this architecture into an operational system involves incorporating value-adding components that include: caching, inter-process communications,
data retrieval from cloud storage, metadata management, data adapters,  and component orchestration.  Figure \ref{ArcaArquitecture} presents the full-blown ArcaDB  architecture. ArcaDB is comprised of six (6)  mayor components: a) Coordination Unit, b) Catalog, c) Tasks Queues, d) Cache Manager, e) Processing Units,  and  f) Cloud Adapter and Storage. In the next sections we shall provide details on each of these components.

%was implemented using a coordinators-workers approach. These two main components are supported by other technologies that makes possible the %execution of queries using SQL language. As shown in Figure \ref{ArcaArquitecture} it's architecture is comprised of six compoments; Coordination Unit, %Catalog, Tasks' Queues, Common Caching Memory, Processing Units and Persistant Storage.

%REVISIT
%
%This coordination algorithm is containerazied, along with Redis and PostgresSQL, to facilitated scaling and deployment.  All of its components are conteainared execepot for Common cache and Persistant storage.
%\begin{figure}[htb]
%	\centering
%	\includegraphics[width=0.42\textwidth]{fig/System ArchitectureV2Poster.png}
%	\caption{ArcaDB Architecture}
%	\label{ArcaArquitecture}
%\end{figure}

\subsection{Coordination Unit}

%Explain what it does without implementation details.

The coordination unit receives queries from clients through a REST API. Upon query  arrival, the coordinator fetches metadata from the catalog, to parse and validate the query. After that, an initial query plan is generated and optimized using a SystemR-style optimizer \cite{Griffiths13}. This plan details the order in which the operations and \acrshort{udf}s should execute. The coordinator then traverses this plan on a  node by node basis with three main purposes: 1) to identify  the implementation of the operation/\acrshort{udf} that is needed, 2) decide the instance types in which the operators should execute, and 3) split the query plan operator into a given number of executable tasks based on factors such as the number of partitions or data type characteristics. Once these three steps are repeated for each query plan node, then the query is ready to be executed.

\begin{figure}[htbp]
	\centering
	\includegraphics[width=\columnwidth]{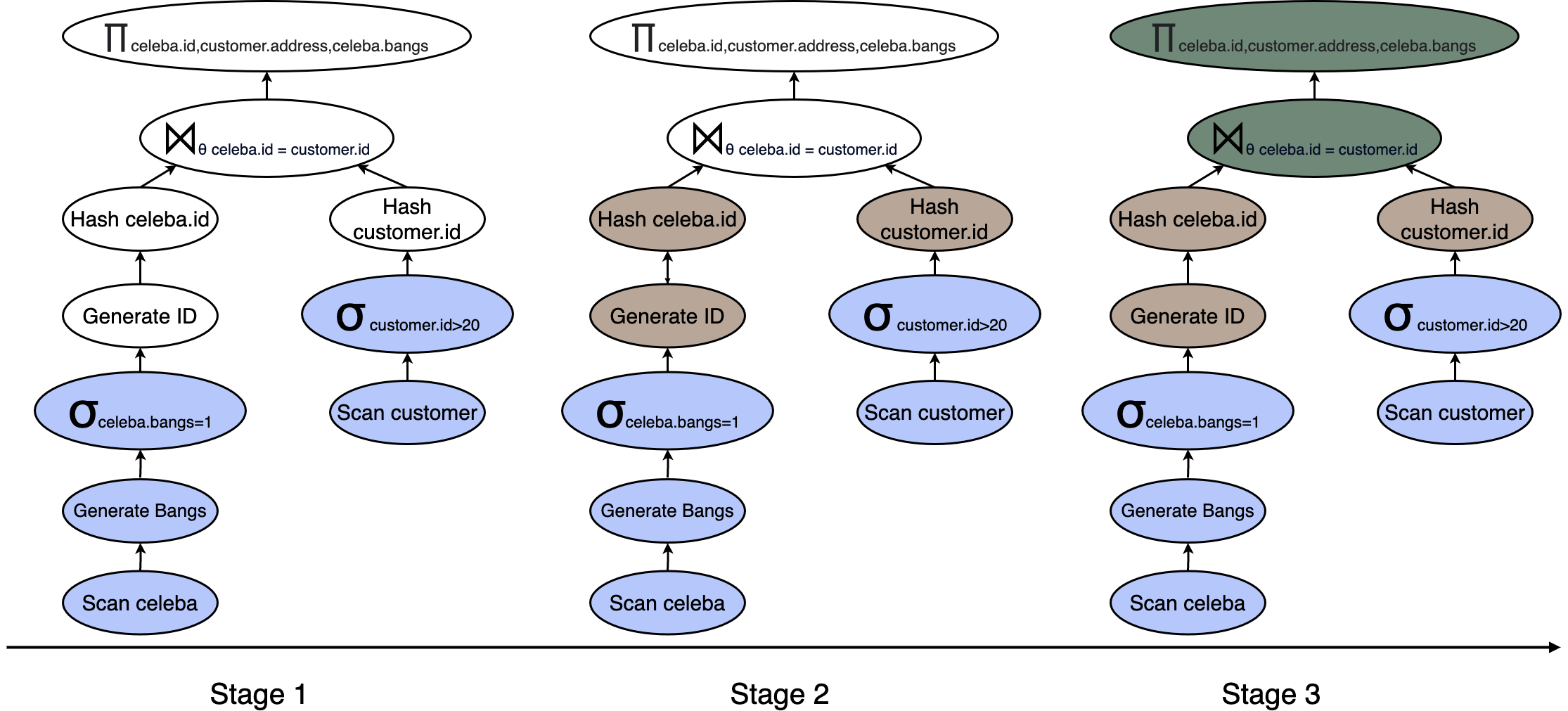}
	\caption{Parallel Query Processing in Stages}
	\label{parallel-plan}
\end{figure}

\par The plan execution involves sending query tasks to the tasks manager, where they are queued based on their resource assignment.  The plan is executed starting at the bottom and through to the top as is shown in Figure \ref{parallel-plan}. In the first stage, scan and selections are run in parallel on various workers. In the second stage, the hash partitioning in performed. In the third and final stage, join probing and result projection are executed. Intermediate results are pipelined  over an in-memory shared cache. Notice that this approach is a departure from traditional top-down query iterators. With this approach, we can release resources and workers as soon as the task they are running is finished. To accomplish this, the coordinator keeps track of which query tasks  have finished and which new tasks should be allocated to execute next. It repeats this process until all nodes in the plan have been executed and query results are placed in the cached memory for the user to fetch.
%Finally the result are send to the user.

\subsection{Catalog}
The catalog receives the query for validation and plan development. It consist of virtual tables and attributes that represent the data stored in a data lake.
In ArcaDB, all data is stored in a data lake platform \cite{10.14778/3415478.3415560, 10.1145/3035918.3056100, 10.14778/3574245.3574274}. With this information the catalog  provides enough metadata for the   optimizer to develop the query plan.
% a plan and optimizes it using System R optimizer. Then the final plan is send to the coordinator.

\subsection{Query Tasks Queues \label{Tasks-Queues}}
\begin{figure}[h]
	\centering
	\includegraphics[width=0.38\textwidth]{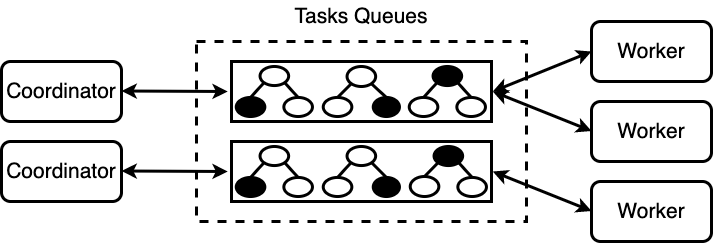}
	\caption{Publish/Subscribe Tasks Broker Model}
	\label{message-broker}
\end{figure}
The Query Tasks Queues take the role of message broker between the coordinator and workers instances, as shown in Figure \ref{message-broker}.
In this case, a Publish/Subscribe model is used, in which each instance is able to subscribe and publish to a specific message queue. Each queue is designed to communicate the coordinator with instances of  specific computational capabilities. For example, the GPU-Queue enables communication with instances that have one or more GPU accelerators. In ArcaDB,  we used this method as the principal communication channel through which tasks are shared.

\subsection{Cache Manager}
As shown in Figure \ref{message-broker}, workers obtain their task from specific queues.
%This happens because workers need to publish query results to  cache store. Also, some of the tasks allocated by %the coordinator require partial  execution thru various stages. An example case is that of a hash join that feeds from %an UDF selection. In this case, all the selection results are needed before the partitioning phase completes.
A task may depend on the completion of another to start. This is why data processing within a disaggregated system is better looked at as a pipeline, where the results of an operation are fed into the next operation. These intermediate result are shared in ArcaDB using a common caching system. An example of this is a hash join operation. The partitioning phase needs the data from the previous operations/UDFs to start. The caching system in this case will hold these intermediate results and makes them available as soon as generated. It also serves to save final results. Ideally, it operates close to the workers so that the network latency has minimal impact on query response time.

\subsection{Query Processing Units}
The query processing units take care of the execution of  query operations such as \emph{selections},  \emph{projections}, \emph{joins} and UDFs.
%Their support to all or some of them, may depends on the resources that they are allocated to
In ArcaDB, operators and UDFs are disaggregated.  Each operator can be run {\em individually} inside one or more containers. Moreover, adjacent operators in the query plan that share the same computational requirements are {\em collocated} in the same container to optimize resource usage and avoid data shipping.
%Together with the cache and tasks queues, workers are able to carry out a plan and hence results to a query. This units provide operations like selections, projections, joins and UDFs. Their support to all or some of them, may depends on the resources that they are allocated to.

The Query Processing Units, as shown in Figure \ref{ArcaArquitecture}, are worker nodes that are containerized. The nodes are the computational instances   that are available within the underlying cloud infrastructure. These instances were shown  in Figure \ref{modernclouds} of section \ref{hetclouds}. ArcaDB supports the usage of these heterogeneous  resources through containers. These portable environments are configured to support specific query operators, which are then allocated to the nodes best equipped  to execute them. Each container is  subscribed to specific queues that  are compatible with the operators or UDFs implemented in their environment.  Given a query, it might be the case that operators are implemented in different programming languages, and used seamlessly in the same plan. For example, we could have a predicate  that is implemented as a PyTorch classifier, and a hash join implemented with Java or C++.
This feature enables the use of specific software infrastructure to define operators.

Any given node might host containers that have similar performance needs but implement different operator/UDFs. For example,  a GPU node might be hosting one container that runs an image classifier, and another container that runs  a text classifier.
 The containerization of operators/UDF, often occurs with specific UDTs, related  operations over entities, or basic data processing functions such as storage, updates or deletions.

\subsection{Cloud Adapter and Storage}
Whether the data is saved locally or stored in the cloud, ArcaDB uses an adapter that standardizes the operations for  retrieval and storage of data.
This module provides the flexibility to support data retrieval of various storage mediums.
% without singularly implementing  adapter for each one. This also allows for usage of various storage systems.

%% file: implementation.tex
\section{ArcaDB Implementation\label{implementation}}
ArcaDB is a  distributed system that relies upon many technologies. It was implemented in approximately 4,000 lines of code.
%This section we details each component implementation.

\subsection{Major Components}
 %Although all components of ArcaDB exisit on a contianerized format, two out of the seven component on the architecture are not conteinerized. In this we explain which technologies were used for each component and its format.
 We implemented our current prototype, with the  architecture mentioned in Section \ref{architecture}, as follows.
{\setlength{\parindent}{0pt}
\begin{itemize}[leftmargin=0.4cm]
	\item  \textbf{Coordinator} - The coordinator is a multi-threaded Java processes, implemented with Java 17. It runs inside a Maven  Docker  container.
	\item  \textbf{Catalog} - The catalog is implemented with a combination of PostgreSQL and JSON files.  We run PostgreSQL inside a Docker container, and use it to hold metadata about virtual tables and their attributes. In addition, each data lake holds JSON files that store the mapping between virtual
	tables and local table files. Also, there are JSON files that specify which schema mapping functions, UDFs and UDTs apply to each virtual table.
	\item  \textbf{Task Queues} - The task queues are implemented with a containerized instance of Redis, an in-memory data storage and  message broker.
	 %is a containerized intance of Redis. It provide the methods to implmented various basic data structures.
	 Although its principal usage is managing task queues, using \acrshort{fifo} as its scheme, it also works as a lookup table for the path of cached files, to  off-load this task from the cache manager.  %for the caching manager.
	 This is because the \emph{list-path} function in the caching system is more costly in comparison with Redis lookups.
	\item  \textbf{Processing Units} - All processing units are containerized. We have implemented units for : a) scanning ORC files , b) selection, c) projections, d) hash joins, and e) UDFs for image classification and scheme mapping. The first four operations are implemented in Java  17 as single thread tasks.  The hash joins are run on high memory nodes, and the other three run on general purpose nodes.
	The UDFs are implemented as Python processes, some of which  use PyTorch for inference. These operators are run on GPU-accelerated instances.
	 This shows that ArcaDB can support heterogeneous implementation of query operators to capitalize on the application-specific libraries that provide best in class performance.

	%They are configured to run on a predefined node type and dequeue query plan fragments (tasks) from a specific queue. We implemented two processing units. On of them suppor where developed. One in Java 17 to support ORC file operations and another in Python 3.8 with Pytorch to support image inferences.
	%They each are hosted by general purpose intances and accelerated instances respectively.
	\item  \textbf{Cloud Adapter and Storage} - We used Alluxios as the adapter to manage data from cloud storage (e.g., S3).
	%The way ArcaDB access persistant data component is through Alluxios. It uses its local HDFS instance to save the tables and all its files.
	\item  \textbf{Cache Manager} - The caching system consists of another instance of Alluxios configured to run on top of high-memory nodes.% with high memory capacity.
\end{itemize}
}

\subsection{Deployment and Orchestration}

In our prototype, we used Docker Swarm as the orchestration tool for containers. This tool makes it easy to define, allocate and scale containers. It provides features like labeling nodes and establishing deployment constrains that match container deployment to specific labeled nodes. For example, if a service is constrained to work on a node with “processingUnit=GPU” then, only nodes that are labeled as GPU, will be able to run those containers. Swarm enforces these services constrains.  If they are not satisfied, containers  are not provisioned.
% This means that each node  joined to the swarm should be properly labeled and constrains have to be correctly defined otherwise, services might be missing.

%\subsection{Implementation Details}
%
%\par As of this prototype, it is asumed that tables are not distributed among different storage units. Persistant Storage and Common Caching memory are components within ArcaDB that are not containerazied. This is due to the nature of their function within the arquitecture. The idea is that they are nodes that solely focus on data management.
%
%THis whould go into THesis but not here.
%
%Processing units are "subscribed"  to serve a specific queue. This is configurable through enviroment variables. An important details is that the containers have to be able to support all operators or \acrshort{udf}s that are queued there. This requires a setup that has to be consonant. The following process details hwo this configuration is achieved
%\begin{enumerate}[label=\arabic*)]
%	\item Define a function and set it to dequeue from specified queue.
%	\item Containerize said function using Docker\cite{merkel2014docker}.
%	\item Modify metadata to allocate functions for specific queue.
%	\item Using Swarm, label nodes based on their resources
%	\item Using Docker compose file,  define a service with it's desired configurations and contrains allocation using swarm node labelling.
%	\item Deploy using docker compose file and Swarm.
%\end{enumerate}

%% file: datamodel.tex
\section{Data Model\label{datamodel}}
ArcaDB follows an object-relational style data model. Data is viewed as a collection of tables called {\em virtual tables}, because they are not necessarily materialized on disk.  Instead, schema mappings and ETL are used to realize the data as needed from their underlying storage location and raw formats. Support for \acrshort{udt}s/\acrshort{udf}s is possible through registration in the system catalog and type declarations in virtual table definitions. 
In this section we detail how the data model works within ArcaDB to deliver SQL language support for datasets that need to be dynamically mapped to an application-specific schema.

\subsection{Virtual Tables \label{virtual-tables}}
Typically, applications in a database system adhere to a schema that is defined before any data are stored. This is known as {\em schema on write} because data items are forced to fit a predefined structure as they are stored. This method is ill-fitted  in situations where data attributes are generated ad hoc via ETL/Schema mapping. This scenario is frequent within data lakes, where data is stored in their raw formats, and might not conform with the application schema. In such situations,  data transformations into the target schema must be performed before any query processing can occur. For this scenario, {\em  schema on read} provides the needed flexibility to fabricate the attributes as the data items are read. 

%To better illustrate this point, let us recall the  IoT application in the health care system from section \ref{usecases}. 
% Consider a fully integrated infrastructure, where patient information is shared among authorized health institutions. Suppose that a patient  gets an X-ray, and the technician uploads it to the data lake.  This will trigger a query that includes  a \acrshort{udf} to determine if the patient has a suspicious screening. This is an example of a derived attribute. Based on this result,  the application can automatically contact 
% the head doctor for further diagnostic. In this example, we see the need to  derive new data  and correlate these items with the patient's health records.

To support SQL-like queries, we need to have a schema describing tables, columns, types, and so on. We adopt a scheme similar to IBM's Virtual Table Interface (VTI). Schema mapping becomes an access method that is invoked to generate the rows or columns as needed,  while also invoking any UDF that is required.

Consider, for example, the CelebFaces (CelebA)\cite{celeba} dataset used to train deep learning models.  A table based on this schema is:

\begin{lstlisting}[language=Python,linewidth=\columnwidth]
 celebea(smiling, young, bangs, recideing_hairline, rosy_cheeks, chubby, bald, eyeglasses, ...)
\end{lstlisting}

 The purpose of this dataset is to learn to extract characteristics from  color photos such as hair color, use of eyeglasses, or age group. 
 The idea is that trained models can then 
process new, \emph{never seen before} images and extract these attributes. The results can then be stored in a table for quick reference, or can be fetched as needed by invoking the UDF.
% Figure \ref{celeba-table} shows the ArcaDB virtual schema for the celebrity data, with attributes that are generated by inference. Notice that full materialization might not be needed if the query is not frequent enough, or if not all attributes are needed.

ArcaDB relies on metadata saved by the catalog to produce schema on read and it is also the component responsible for describing which functions could operate over entities and what UFDs are supported by the system.

%\begin{figure}[htbp]
%	\centering
%	\includegraphics[width=0.16\textwidth]{fig/celeba-table.png}
%	\caption{Virtual Table Definition for CelebA Dataset}
%	\label{celeba-table}
%\end{figure}

\subsection{Access Path for Virtual Tables}

\par As we explained in Section \ref{virtual-tables}, attributes are not realized unless projected by the query or used to compute results (e.g., in a predicate). Returning to the example of the CelebA dataset, 
%now that we have feature that could be enquiered,
 suppose that we have the images distributed across various cloud storage systems.  Now, we need to retrieve image data from all  these partitions to extract the derived attributes required by each query. 
 %The retrival of all the partitions is necesarry to materialized a projectiong pertaining to the query. 
 Since these partitions are not necessarily stored in the same path or storage unit, a catalog is needed to keep track of where all partitions are stored. This provides a direct access path to fetch all of them with a remote-fetch scan operator. 
 It is possible to conduct schema mapping as part of this fetch, or it can be done as a follow up operation. 
 %at this point, or 
 Other access methods include grouping attributes or indexing certain characteristics. For example, we can store all images with subjects that have eyeglasses in the same folder in a cloud service, or create an index to access all the files that contain that particular attribute across cloud services. 

\par Since we can allocate multiple containers of each kind in separate nodes or combine them in a few nodes, we need parallel data retrieval because some query tasks could be divided into shards. 
 %when task are divided. 
 %This is characteristical of a Shared Memory Architecture. 
 Partitioning permits simultaneous data processing in various containers. Data is either naturally partitioned in the sense that there are different access paths, or the coordinator divides the tasks based on the amount of files that constitute the table. 

%% file: queryprocess.tex
\section{Query Plan Generation \label{query-plan-sec}}

\begin{figure}[!h!]
	\centering
	\includegraphics[width=0.30\textwidth]{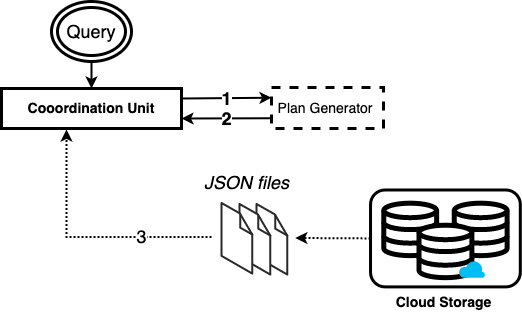}
	\caption{Query Plan Generation}
	\label{plan-generator}
\end{figure}

The plan generation for ArcaDB is divided into two phases: initial plan generation and physical plan generation. This is shown in Figure \ref{plan-generator}. The first phase, labeled with a 1, deals with join order and other operator placement. This is driven by metadata from the catalog and executed by the query optimizer.  The second phase involves assigning resources for each node within the plan. In this phase the coordinator retrieves metadata from the cloud manager to further validate, locate the dataset, and identify available cloud resources to place the operators.
The metadata from the cloud is used  to divide tasks into batches based on number of partitions of the data.
Then, the metadata is used to heuristically assigns specific resources to complete execution of nodes. 

%The first phase, 1 and 2 from Figure \ref{plan-generator}, include sending the query to the catalog and generating an initial plan using virtual tables and attributes. After the plan is send back to the coordination unit, the second phase involves assigning resources for each node within the plan. In this phase the coordinator retrives metadata from the cloud manager. This data is first used to divide tasks into batches based on number of files.

\subsection{Physical Plan and Dissagregation}
% figure is in background before COntainers section since latex nature forces it to be before.

\setlength{\textfloatsep}{3pt}
\begin{algorithm}
	\caption{Resource Assignment for Tasks in Plan}
	\label{alg:resource-assignation}
	\begin{algorithmic}[1]
		\State $plan \gets physical\ plan$
		\State $node \gets null$
		\For{\texttt{$each\ node\ in\ plan$}}
		\If{$node$ is  structured data}
		\If{$node$ operation is a join}
		\State $node.resources.processing \gets CPU$
		\State $node.resources.memory \gets XL$
		\State $node.resources.disk \gets NVME$
		\ElsIf{$node$ operation is a simple projection \\ \ \ \ \ \ \ \ \ \ \ \ \ \ \ \ or is a simple UDF projection}
		\State $node.resources.processing \gets CPU$
		\State $node.resources.memory \gets M$
		\ElsIf{$node$ operation is a selection or a scan}
		\State $node.resources.processing \gets CPU$
		\State $node.resources.memory \gets L$
		\EndIf
		\ElsIf{$node$ is  complex UDF operation}
		\State $node.resources.processing \gets GPU$
		\State $node.resources.memory \gets L$
		\Else
		\State $node.resources.processing \gets CPU$
		\State $node.resources.memory \gets M$
		\EndIf
		\EndFor
	\end{algorithmic}
\end{algorithm}

%\begin{lstlisting}[language=python]
%In: Query Plan
%Out: Null
%for each node in the tree (using any tree traversal)
%if node.datatype == structured:
%	if node.operation == join:
%		node.resurces.processing = CPU
%		node.resurces.memory = XL
%		node.resource.disk = NVE
%	else if node.operation == projection:
%		node.resurces.processing = CPU
%		node.resurces.memory = M
%	else if node.operation == scan |
%	node.operation == selection:
%		node.resurces.processing = CPU
%		node.resurces.memory = L
%else:
%	node.resurces.processing = GPU
%	node.resurces.memory = L
%
%\end{lstlisting}
\par When ArcaDB recieves a query, a plan is generated based on a System-R optimizer. This plan is later transformed 
%translated into a format that the system understands. This include transforming the plan 
into an executable tree data structure that contains all the details for each task. These details include process dependencies, sources from which data should be extracted, data formats, partitions and the operation to be executed. This information is provided in each node. 

The {\bf Algorithm \ref{alg:resource-assignation}} shown above describes the heuristic approach in which the coordinator assigns resources to the plan within ArcaDB. This approach  is prefered mainly due to the dynamic nature of resources. Cost-based alternatives typically assume static allocation of resources, which is not the case for cloud analytics engines.
In our current prototype, we assume that our datasets are images or alphanumeric relational data. The physical plan produced by the optimizer is further modified by the coordinator. The algorithm checks each node in the plan and based on the data types and operators  determines which resources to assign.  If  the operator handles structured data, then further characteristics are considered. Otherwise, image data will be handled, and  the resources allocated for that particular node are accelerators with high memory capability. The operations that were implemented for structured data were {\em join}, {\em projection}, {\em selection} and {\em scan}. If the tree node has a join operation, it is  allocated to resources with very large memory and fast disk. If it has a projection, they are allocated to resources with average memory capabilities. Lastly, if the tree node has a selection or a scan they are assigned resources with large memory. The coordinator does these assignments so that each task has specific sites in which they execute once the plan is disaggregated and dispatched. As part of our on-going work we are exploring reinforcement learning methods to improve on Algorithm \ref{alg:resource-assignation} for more complex cases with dynamic resource availability and fluctuating network conditions within the cloud.  

In figure \ref{fig:query-dissagregation3} we saw how a query is disaggregated from its tree structure. Notice that tasks (i.e., plan nodes) are seen as individual units and the coordinator makes sure to allocate each task to its respective queue. Although not shown in the picture, tasks that rely on other(s) or blocking operations are either managed by the coordinator or automatically allocated after any  underling tasks are finished.

\subsection{Single Table Processing for UDT \label{query-single-table-sec}}

\par Operators and \acrshort{udf} are defined by the users,  and provide the means to extend ArcaDB to support complex functions on \acrshort{udt}. This means that even for structured data, operators could be redefined by the user. Let us take as example the following query:

\begin{lstlisting}[language=SQL, linewidth=\columnwidth]
   select a.id, from celeba as a where hasBangs(a.id)
\end{lstlisting}

\noindent In this query,  the table \emph{celeba} is read using  a schema on read approach. The data is stored in JPEG binary files in a directory in the data lake. The table has an inferable attribute called \emph{bangs}, which will be used to filter the results. This attribute is extracted from a UDF that implements a machine learning classifier.
The other attribute, called {\em id}, is extracted from the image filename.

\begin{figure}[h]
	\centering
	\includegraphics[width=0.3\textwidth]{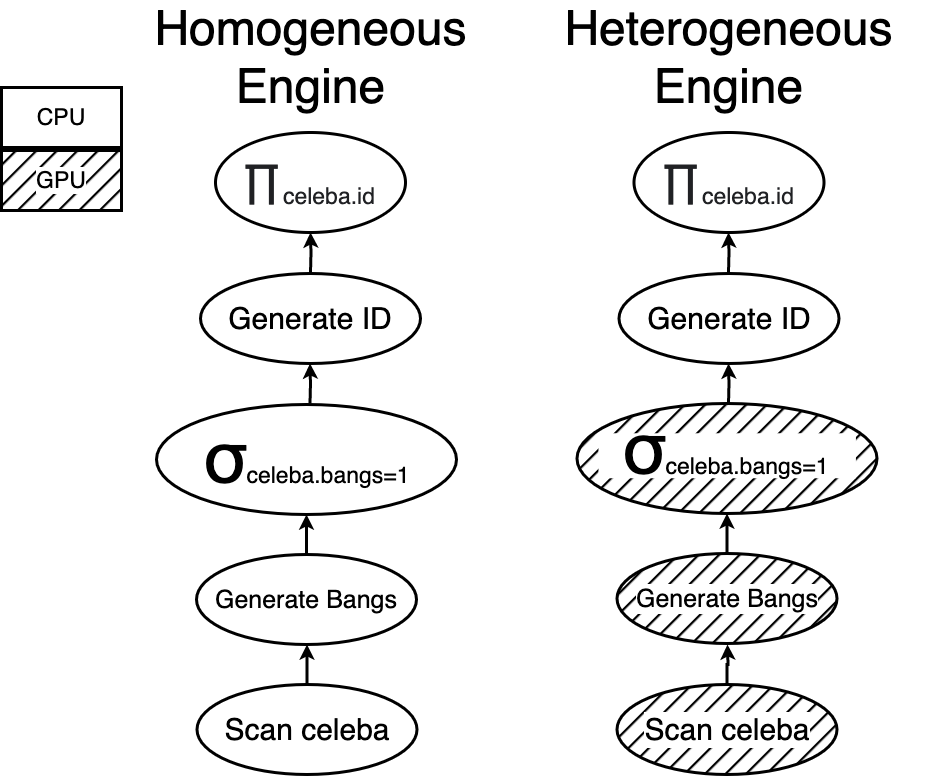}
	\caption{UDT Physical Query Plan}
	\label{image-query-plan}
\end{figure}

\par Figure \ref{image-query-plan} details the plan generated to satisfy the query shown above. Notice that there is scan node to read the images,
a schema mapping node (denoted as Generate T, where T is the inferable attribute) where the attributes are derived, and then we have the standard selection and projection operations.
On the left hand side of the figure, we have the symmetric plan as would be used in an homogeneous engine, which will execute completely in CPU (i.e., no accelerators).  
On the right hand we have the disaggregated plan for ArcaDB engines, which allocates the tasks to different resources. The shaded nodes are assigned to GPU instances.
 %while the heterogenous allocates tasks to different resources.

The inference of the \emph{bangs} attribute and its filtering are best fitted to execute in accelerators, therefore both are allocated to instances with a GPU.
Although they have separate nodes in the plan, these two are executed in the same worker unit, without incurring in the cost of data exchange: the image is retrieved, the bangs attribute  is inferred with a classifier, and then the filtering happens. If there are various attributes to be inferred and filtered, ArcaDB allocates their operations to the same unit if it has support to infer or operate over them. This has the advantage of maximizing resource usage, while avoiding data exchange or multiple data retrival.
 %and filtering conditions that run in the same node, depending on the \acrshort{udf}, another task can be created or all attributes are sequentially <----- ESA ES LA PALABRA? inferred in that same node.
The cloud adapter can mitigate this action if there is a caching system. Although it is preferred that each file is retrieved once per query. For the specified query example, the \emph{id} calculation is done on a separate task on  a general purpose instance, and can be collocated with the projection operation. Considering this information, the plan in Figure \ref{image-query-plan} can be minimally run in two nodes. Image scan, bangs inference,  and \emph{selection} are executed together in a GPU node. Then another task is generated to calculate the {\em id} with a UDF, and project it. In practice, multiple workers running in multiples nodes of types {\em GPU} and {\em general purpose} will be used to run this query.
 %This generation uses another \acrshort{udf} that is deployed to a different resource.

\begin{figure}[h]
	\centering
	\includegraphics[width=0.42\textwidth]{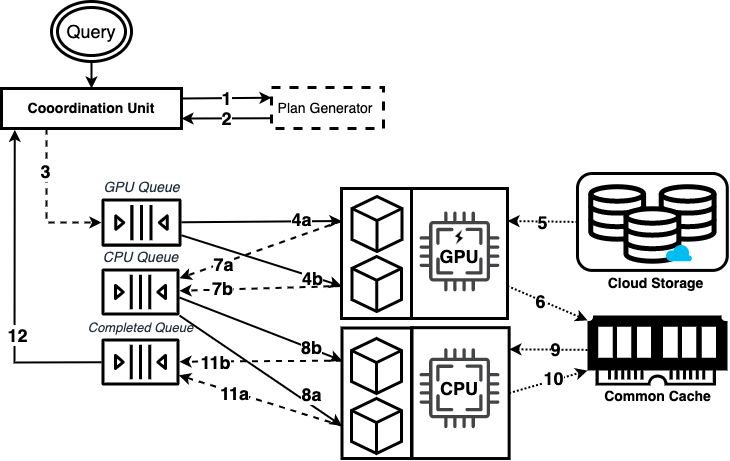}
	\caption{UDT Query Process Flow}
	\label{non-structured-data-query-single-flow}
\end{figure}

Putting it all together, in Figure \ref{non-structured-data-query-single-flow} we describe the process followed by ArcaDB to resolve the query presented in this section. First the SQL query is sent to the plan generator (1) for the first phase. After validation and optimization (2) a plan is returned to the coordinator. Then, the second phase of query generation takes place. Query tasks are divided and assigned to resources.  Query tasks are queued (3) into the task broker and later (4) dequeued by working instances. For this query in particular, a GPU node (5) retrieves data from cloud storage to make \emph{bangs} inferences and filter results. After this process finishes, (6) results are saved to the common cache and (7) tasks for computing and projecting  the \emph{id}  attribute are allocated to CPU instances. These instances (8) receive the tasks and (9) pull previous results from cache memory. Later, they project the resulting columns on the selected files and (10) save them again to the cache. As worker instances finish, (11) they report back to the coordinator and the later (12) concludes plan execution after receiving all completed tasks.

\subsection{Join Processing}

In the presence of \acrshort{udt},  schema mapping operations can be used to transform the data into the UDT format, and the result can be used
in selections, projections, and joins.
The following query shows how this can be achieved.

\begin{lstlisting}[language=SQL]
	select a.id, a.bangs, b.address
	from celeba as a inner join customer as b  
	on(a.id=b.id)
	where b.id>20 and hasBangs(a.id);
\end{lstlisting}

In this query the tables \emph{celeba} and \emph{customer} are joined using \emph{id} as the joining column. Since \emph{celeba} schema is realized on read, the attributes \emph{id} and \emph{bangs} need to be inferred and calculated respectively. On the other hand, \emph{Customer} needs \emph{address} and  \emph{id} to be projected. In this case, these two columns exists in the underlying source. The  filter customer.id  > 20 is applied but as explained in section \ref{query-single-table-sec} it can be  applied as data is retrieved  in the scan operator.
% by their respective operators or \acrshort{udf}.

\begin{figure}[h]
	\centering
	\includegraphics[width=0.4\textwidth]{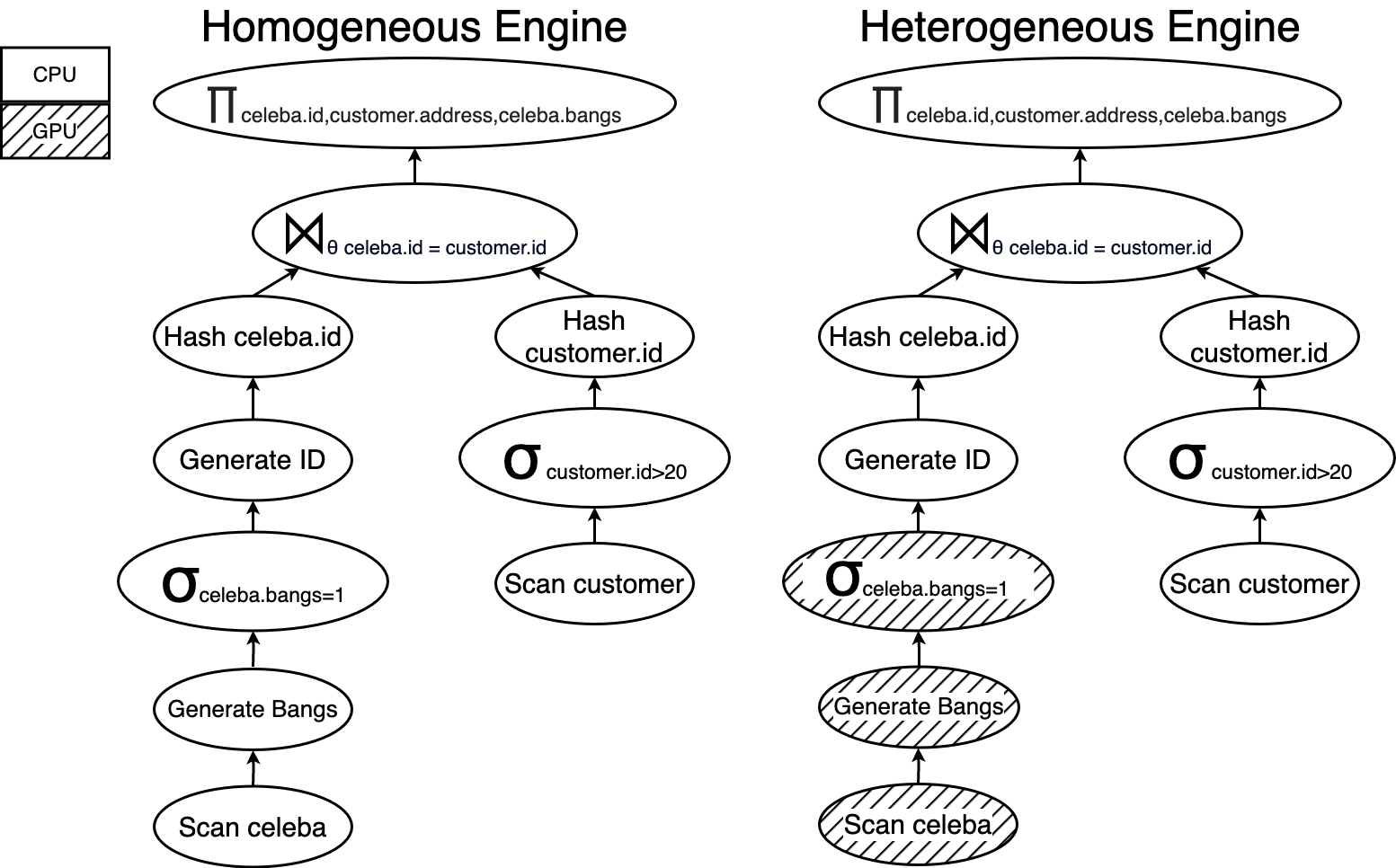}
	\caption{Join Physical Query Plan}
	\label{image-customer-plan}
\end{figure}

Figure \ref{image-customer-plan} shows the plans to execute the query discussed above.  As before, the left-hand side show the symmetric query plan
for homogeneous environments. The right-hand side shows the disaggregated query plan that ArcDB uses.
Similar to the case from the previous section, the second plan allocates tasks related to inference in table \emph{celeba} to nodes with GPUs.
Since data in table \emph{customer} has a defined schema is treated as alphanumeric relational data and allocated to CPU nodes. Even though one table follows a more complex schema on read, (\emph{celeba}), and the other follows a direct mapping from the schema, (\emph{customer}), the join is possible because the attributes are realized as needed, in an ad-hoc fashion before the join is done.

\begin{figure}[htb]
	\centering
	\includegraphics[width=0.44\textwidth]{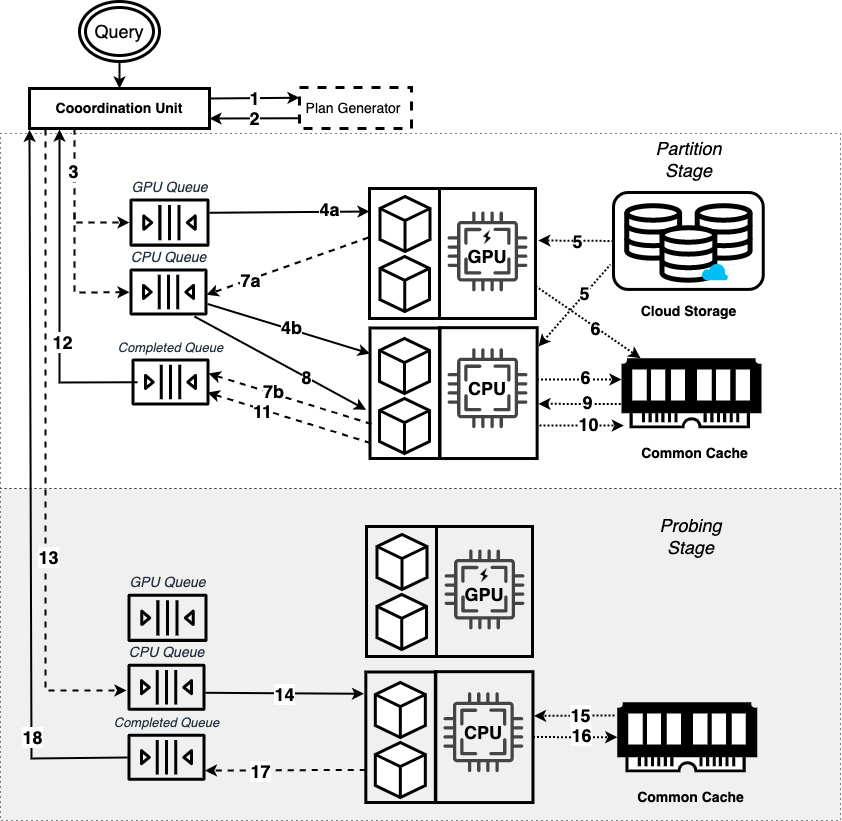}
	\caption{Join Query Process Flow}
	\label{non-structured-data-query-join-flow}
\end{figure}

Figure \ref{non-structured-data-query-join-flow} illustrates the process to deliver the results for the join query discussed in this section. The join algorithm used for this example is a distributed version of the GRACE Hash-Join \cite{10.1007/BF03037022}. The algorithm is composed of two stages. The first is partitioning, which involves hashing and separating data into buckets. The second phase is probing, in which respective buckets from each table are joined. All the buckets are copied to the cache, which makes them available to other container instances.  Thus, the shuffling phase is replaced by reading from the cache.

Putting it all together,  let us retrace the steps in Figure \ref{non-structured-data-query-join-flow} . The plan is generated and tasks are assigned their respective resources (1-3). Tasks are simultaneously being dequeued (4), and each container instance pulls its own and then retrieves the data as needed (5). After execution of schema mapping, and selections the results are saved to cache memory (6). The next step (7) varies from instance to instance.
In (7a) the worker running in the GPUs sends a message to the CPU queue to signal that the partitioning phase can begin. This triggers a new task in one or more CPU nodes for  containers to execute the partitioning phase. In 7(b) the selection of customer data is completed and the container in the CPU node reports back to the coordinator with the concluded tasks.

Following the example in the figure, 8) CPU instances take on the task of producing and hashing the attribute \emph{id} from table \emph{celeba}. Since accelerated instances already filtered files, 9) computing instances pull previous results from cache memory. After calculations and hashing, 10) these instances save their results in cache again and 11) report back to the coordination unit. 12) The coordinator then waits for all hashing to finish and this concludes the join partitioning stage. Once again, 13) the coordinator assigns tasks that contain the buckets to be joined. 14) These tasks are taken on by the computing instances and 15) a pair of matching buckets from both tables is retrieved from cache memory for joining. 16) Results are once again saved to cache memory and 17) instances report back their finished tasks. To conclude the plan execution, 18) the coordinator waits for all buckets to be joined. After the probing phase is done, the query plan execution is concluded.

%% file: performance.tex
\section{Performance Evaluation\label{performance}}
The main objective of this preliminary performance study is to  illustrate the substantial performance benefits provided by a system that uses a disaggregated engine to exploit accelerators, such as ArcaDB. We contrast our ArcaDB architecture  with  one that lacks this capability, and must rely heavily on parallelized tasks among CPU-based workers.

\subsection{Setup}

This section details the performance metrics and platform used to obtain results.

\subsubsection{Performance Metrics}
To evaluate the effectiveness of the ArcaDB architecture the following metrics were collected.
\begin{itemize}[leftmargin=0.4cm]	
	\item \textbf{Response Time:} This metric measures the total wall time that it takes to complete a query.
	\item \textbf{Instance Cost:} We estimate  the cloud instance cost to run each query, using different prices from cloud services, query  response time, and the number of allocated instances. 
\end{itemize}

\subsubsection{Cloud Platform}
{\small 
\begin{table}[!h!]
	\begin{center}
		\caption{VMs in Cluster Specifications\label{tableVMSpects}}
		\begin{tabular}{p{2cm}|c|c|c|c} % <-- Alignments: 1st column left, 2nd middle and 3rd right, with vertical lines in between
			\centering \textbf{Usage} & \textbf{\#} & \textbf{Processing Unit} & \textbf{RAM} & \textbf{Disk}\\
			\hline
			\centering Workers, manager, and catalog&12&8vCPU&64GB & 100GB\\ \hline
			\centering Virtual Cloud Storage & 4 & 16vCPU & 128GB & 200GB\\ \hline
			\centering Cache & 2 & 24vCPU & 181.6GB & 500GB \\ \hline
			\centering GPU Workers & 2 & 24vCPU + 1GPU & 87.9GB & 250GB \\
		\end{tabular}
	\end{center}

\end{table}
}

\par ArcaDB was implemented on a private cloud built with OpenStack, using 20 \acrshort{vm}s as specified  in Table \ref{tableVMSpects}. Services within the engine are distributed among the available resources taking into consideration possible bottlenecks and memory contention. Workers nodes are VMs in the cloud that are solely focused on serving as hosts for worker containers that perform task execution. The array of workers was modified as shown by each experimental query run. The hardware operates as a {\em private cloud} platform that provides these VMs  and runs Red Hat Open Stack Wallaby. The underlying hardware is as follows: a) 10 high-memory Compute Nodes, and b) 12 GPU Nodes. Each Compute Node has 2 x  12-Core Intel Xeon 3.7 GHz CPU, with 384 GB of RAM, and 2 TB HDD. Each GPU Node has  1x  12-Core Intel Xeon 3.7 GHz CPU, with 192 GB of RAM, 1x NVIDIA Tesla V100s 32 GB GPU, and 4 TB HDD. All nodes are connected to a 10Gbps Ethernet switch.

\subsubsection{Software Platform}
The bare metal nodes that form the cloud run Linux CentOS 7. All VMs run Ubuntu Linux 20.04 LTS. Most components of ArcaDB are implemented in Java 17. UDFs are implemented in Python 3.8 and PyTorch 1.13 with GPU support enabled for NVIDIA CUDA 11.7. The catalog was implemented with PostgreSQL 14. The message queues were implemented with Redis 6. All containers are implemented and managed with Docker 20.10. All containers  for the coordinator and workers are run on a modified version of the UNIX alpine container.
The containers for UDF run on a modified version of the Ubuntu 20.04 container. The cache  in the cluster and the cloud adapter were run on Alluxios 2.8.

%Since the catalog is access only at the beginning of each query request, it shares resources with the implementation of tasks queues. Virtual Cloud instanes and Cache Common System are two separated groups of computers. These computers run Alluxios respectiveley ewith two different configurations. Since they are frequently accessed by worker nodes and to avoid bottlenecks when reading persistent data and caching intermediate results, this two are treated separately and configured as so. The frequency depends of the query and plan.  It is compose of VMs with high memory capabilities. We had the data stored locally in the cluster. Emulating havins all the services in the same cloud service.

%\subsubsection{Data Sources - ORC files}
%%Unlike typical query engines,
%ArcaDB stores structured data using columnar format, \acrfull{orc}.This format is suited for \acrshort{olap}, where queries often operate in a short selection of columns. Unlike row formats, tuples can be partially retrieved without reading them entirely. Only projected or needed columns are extracted from the file. For the experiments, Apache ORC files 0.11 were used.
%
%\par \acrshort{orc} file have a default compression which reduces the size depending on the datatype. For TPC-H compression rates are around 75\% per file. In a distributed system this is a desirable trait to avoid network data congestion.

\subsubsection{Benchmark Queries}
\par Currently, there is no well-recognized benchmark by the database community to evaluate read-only workloads of both heterogeneous data and complex UDFs  over clouds.  TPC benchmarks are either
based on homogeneous alphanumeric data, or based on combined queries/AI programs (i.e., TPCX-AI).
For this reason, we designed our experiments employing various known datasets that showcase ArcaDB capabilities. 

To test the prototype for ArcaDB and show its capabilities we developed six queries shown in Table \ref{table:benchmarkqueries}. These queries used three main data sets. 
The first one was TPC-H \cite{tpc-h} with a scale factor of 100. Each table was saved to various ORC files. 
%In order to have an uniformity on task load distribution, ORC file where predefined to have 256MB of data each. Using Pandas framework the Formula \ref{formula:ORCsize} %was used to determine the amount of tuples for each file to satisfy this constrain. 
Each ORC file was saved under a folder in the data lake with its respective table name.

%\begin{equation}
%	\label{formula:ORCsize}
%	Tuples\ per\ file=\frac{256*1024^2}{\frac{{\sum_{i=1}^{n=10} randomTupleSizeFromTable()}}{10}}
%\end{equation}

\par Currently, there is no dataset to benchmark UDF in non-structured data. Therefore, the experiments use CelebA \cite{celeba} and PubChem \cite{10.1093/nar/gkac956} as a representation of non-structured data. The first has approximately 200k images with 42 classified attributes and roughly 2GB worth of JPEG images. The second is a 42MB CSV file that contains string representations of molecular structures, their names and other identifiers. These files were saved locally to the Virtual Table Interface (VTI) under folders named "celeba" and "pubchem" respectively. A classifier for hair bangs and a regression model for  molecular weight were trained on the data and added to ArcaDB as a \acrshort{udf} to make inferences.

\begin{table}[h]
	\begin{center}
		\caption{Benchmark Queries \label{table:benchmarkqueries}}
		\begin{adjustbox}{max width=\columnwidth}
		\begin{tabular}{|l|l|} % <-- Alignments: 1st column left, 2nd middle and 3rd right, with vertical lines in between
			\hline
			\centering \textbf{Query \#1 (\,Q1)\,} & \textbf{Query \#2 (\,Q2)\,}\\
			\hline
			\begin{lstlisting}[language=SQL, basicstyle=\small]
select id, hasEyeglasses(a.id),
hasBangs(a.id)
from celeba as a
			\end{lstlisting} &
			\begin{lstlisting}[language=SQL, basicstyle=\small]
select id, smile, isometric,
molecular_weight(id) as weight
from PubChem
			\end{lstlisting} \\
		\hline
		\centering \textbf{Query \#3 (\,Q3)\,} & \textbf{Query \#4 (\,Q4)\,}\\
		\hline
			\begin{lstlisting}[language=SQL, basicstyle=\small]
select * from celeba as a
where hasEyeglasses(a.id)
and hasBangs(a.id)
			\end{lstlisting} &
			\begin{lstlisting}[language=SQL, basicstyle=\small]
select id, smile, isometric,
molecular_weight(id) as weight
from PubChem
where molecular_weight(id) > 437.9
			\end{lstlisting} \\
		\hline
		\centering \textbf{Query \#5 (\,Q5)\,} & \textbf{Query \#6 (\,Q6)\,}\\
		\hline
			\begin{lstlisting}[language=SQL, basicstyle=\small]
select id, smile, isometric,
molecular_weight(id) as weight
from PubChem
where molecular_weight(id) > X 
and exact_mass(id) > 200
			\end{lstlisting} &
			\begin{lstlisting}[language=SQL, basicstyle=\small]
select a.id, b.address,
hasEyeglasses(a.id),
from celeba as a inner join 
customer as b
on(a.id=b.id)
where b.id>20 and 
hasEyeglasses(a.id);
\end{lstlisting} \\
	\hline
		\end{tabular}
	\end{adjustbox}
	\end{center}
\end{table}

\subsubsection{Cloud Instance Prices}
\par Instance cost analysis was based on Amazon EC2 prices. We matched up our Instances  with similar EC2 specifications and prices were selected using "On-Demand." These instances pay by minutes of usage, rounding up, and assuming all computers are engage thought the query execution. For {\em memory intensive} nodes, we took \emph{rad.2xlarge} as reference, which cost \$0.524 per hour and divides to \$0.0087 per minute. For {\em GPU accelerated instances}, we took \emph{p3.2xlarge} as reference, which cost \$3.06 per hour and divides to \$0.051 per minute. 

%Table \ref{table:AmazonCosts} details the instance and cost to execute those analysis.
%
%\begin{table}[!h!]
%	\begin{center}
%		\caption{EC2 Instance Price\label{table:AmazonCosts}}
%		\begin{tabular}{ccc} % <-- Alignments: 1st column left, 2nd middle and 3rd right, with vertical lines in between
%			\centering \textbf{Intance} & \textbf{Price per Hour} & \textbf{Price per Minute} \\
%			\hline
%			rad.2xlarge & \$0.524 & \$0.0087 \\
%			p3.2xlarge &  \$3.06  & \$0.051  \\
%		\end{tabular}
%\end{center}
%\end{table}

%A cost analysis using Amazon EC2 as reference and matching up the instance to the closest offered but amazon.
%Taking into consideration the time it takes and using EC2 as reference as to how much this query might cost to run in the cloud.
%r5ad.2xlarge or r5.2xlarge= (with similar capacity to our CPU workers) cost \$0.524 or \$0.576 per hour per intance.
%p3.2xlarge cost =  \$3.06

%\par Currently, there is no dataset to benchmark UDF in non-structured data. Therefore, the experiments use CelebA \cite{celeba} and PubChem \cite{10.1093/nar/gkac956} as a representation of non-structured data. The first has approximately 200k images with 42 classified attributes and the second is a 42MB CSV file that contains string representations of molecular structures, their names and other identifiers. These files were saved locally to the Virtual Table Interface (VTI) under folders named "celeba" and "pubchem" respectively. A classifier for hair bangs and a regression for  molecular weight were trained using a subset and added to ArcaDB's as a \acrshort{udf} to make inferences.
\subsection{Test Configurations}

For  all  queries, in the plot for response time, the  x-axis indicates  system configuration, and the y-axis provides the response time in minutes. All queries were run three (3) times, and the average was taken and reported in the plots.
An apple to apple comparison to others popular analytic engines is not suitable since they dont share a common baseline. Instead we ran queries in various  configurations: a) non-disaggregated with one worker that uses only CPU (e.g., a single node configuration) , b) non-disaggregated with various workers that use only CPU (e.g., as shared nothing system - for computational resources only),   c) disaggregated engine with one worker that has a GPU and one worker that has a CPU, and d) disaggregated engine with several workers that have  CPUs and GPUs. Configurations (c) and (d) are the native ArcaDB configurations.
  For the sake of presentation simplicity, we shall label non-disaggregated configurations as {\em CPU}, and disaggregated configurations as {\em GPU}. Notice, however, that in ArcaDB other types of instances and accelerators could be used. 
\subsection{Generalized Projection Queries}
\subsubsection{Query \#1}
%\begin{lstlisting}[language=SQL]
%	select id, hasEyeglasses(a.id),
%	hasBangs(a.id)
%	from celeba as a
%\end{lstlisting}

\begin{figure}
	\centering
	\subfloat[Response Time]{
		\includegraphics[width=0.35\textwidth]{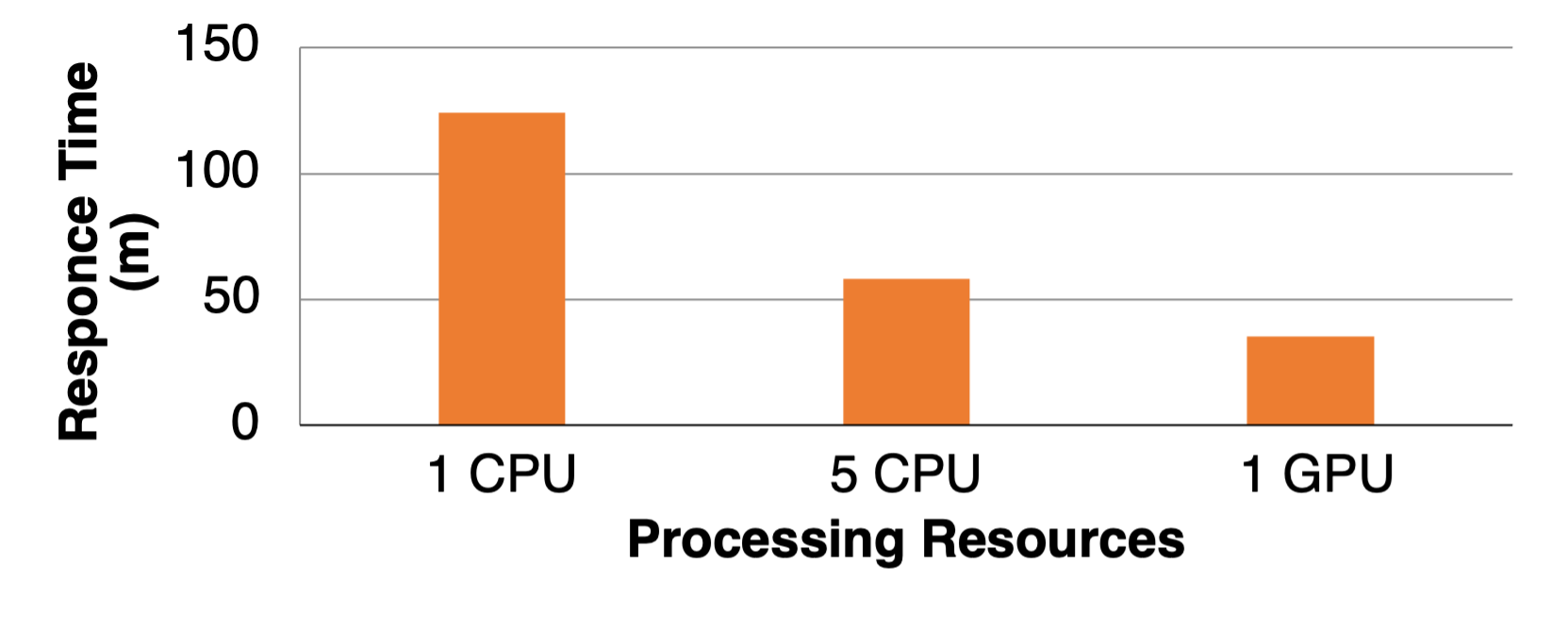}
		\label{query1-results}
	}
	\subfigurespace
	\subfloat[Cost Calculations]{\small
		\label{table:amazonEC2calsq1}
		\begin{tabular}{ccccccc} % <-- Alignments: 1st column left, 2nd middle and 3rd right, with vertical lines in between
			\\
			\centering \textbf{Case} & \textbf{Min} & \textbf{CPU} & \textbf{\$/min} & \textbf{GPU} & \textbf{\$/min} & \textbf{Total}\\
			\hline
			1 CPU & 125 & 1 & \$1.09 & 0 & \$0 & \$1.09 \\
			5 CPU & 59 & 5 & \$2.57 & 0 & \$0 & \$2.57 \\
			1 GPU & 36 & 0 & \$0 & 1 & \$1.84 & \$1.84 \\
		\end{tabular}
	}
	\caption{Results for Query \#1}
\end{figure}

This query contains two UDF generalized projections that extract features from the image table: 1) if the person has eyeglasses, and 2) if the person has hair bangs.  Both are implemented as PyTorch classifiers. 
Figure \ref{query1-results} shows results for query \#1. Notice that using a GPU and disaggregated organization provides the best results, even beating the configuration with five workers. 
When the  query is run on one high-memory instance, it gets the slowest response time at about 125 minutes. In contrast, using a disaggregate approach and exploiting  a GPU accelerator results in a total query response time  of 36 minutes. Notice that  just 1 GPU makes the query 3.5x faster. Interestingly, the cost-to-response time trade-off for this query is best when run disaggregated.   A comparison using monetary cost of the resources shown in Figure \ref{table:amazonEC2calsq1}. The five node configuration has the highest cost and second-best response time. The disaggregated configuration is cheaper by \$0.73 and has the best response time. 

\subsubsection{Query \#2}

%\begin{lstlisting}[language=SQL]
%	select id, smile, isometric,
%	molecular_weight(id) as weight
%	from PubChem
%\end{lstlisting}

\begin{figure}
	\centering
	\subfloat[Response Time]{
	\includegraphics[width=0.35\textwidth]{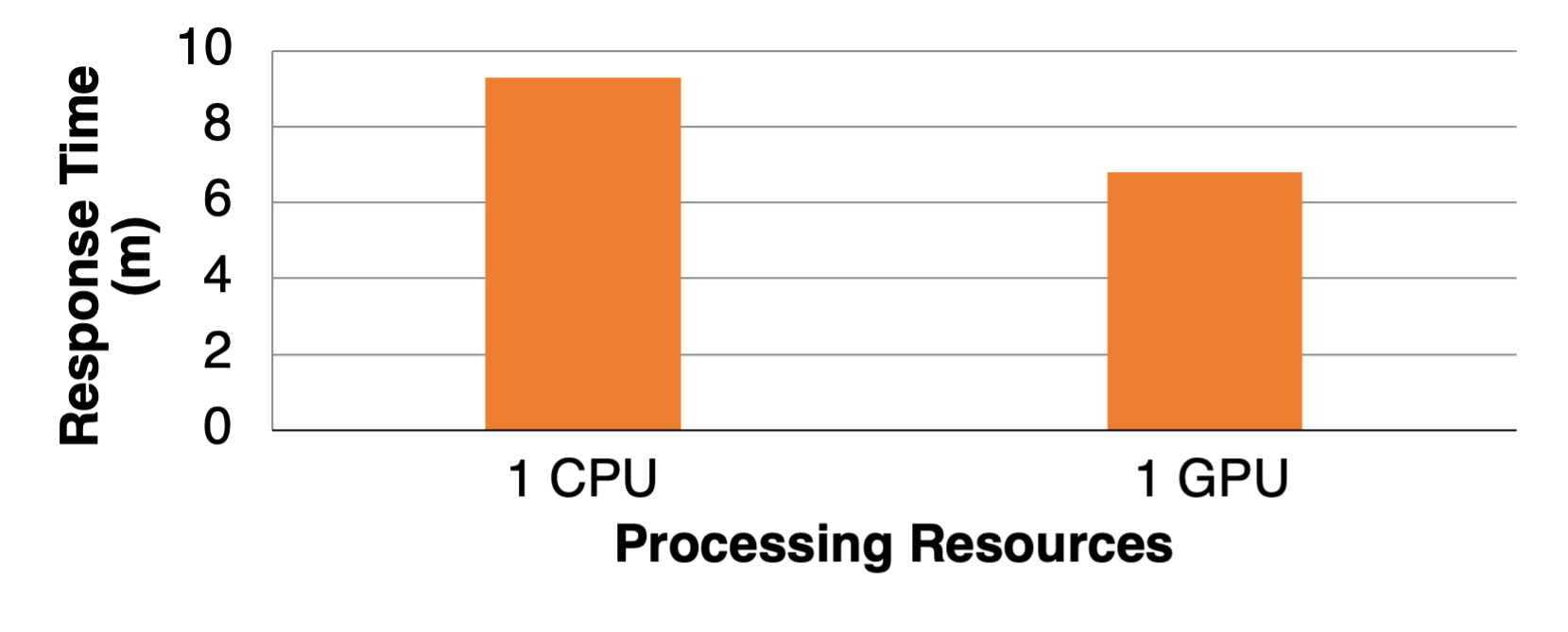}
	\label{query2-results}
}
\subfigurespace
\subfloat[Cost Calculations]{\small
		\label{table:amazonEC2calsq2}
		\begin{tabular}{ccccccc} % <-- Alignments: 1st column left, 2nd middle and 3rd right, with vertical lines in between
			\\
			\centering \textbf{Case} & \textbf{Min} & \textbf{CPU} & \textbf{\$/min} & \textbf{GPU} & \textbf{\$/min} & \textbf{Total}\\
			\hline
			1 CPU & 10 & 1 & \$0.09 & 0 & \$0 & \$0.09 \\
			1 GPU & 7 & 0 & \$0 & 1 & \$0.36 & \$0.36 \\
		\end{tabular}
}
\caption{Results for Query \#2}
\end{figure}
This query has a UDF as a generalized projection that is used to compute the molecular weight of a molecule. This function is also implemented as a PyTorch classifier over strings. 
Figure \ref{query2-results} shows results for query \#2.  This data set has millions of rows, but overall, it was small and fitted in the memory of just one worker. When the  query is run on one high-memory instance, it get the slowest response time at about 10 minutes but when  accelerated it takes about 7 minutes, which is about 1.4x faster. This performance improvement is not as impressive for two reasons: a) the table is small, and b) the UDF is run on a small object and its benefits are not as noticeable. Moreover,  comparison using monetary cost of the resources shows that the CPU is cheaper by \$0.27 which makes the accelerated instance cost almost 300\% more. 
We further expand on this issue in the discussion section. 

\subsection{Selection Queries}

\subsubsection{Query \#3}

%\begin{lstlisting}[language=SQL]
%	select * from celeba as a
%	where hasEyeglasses(a.id)=1
%	and hasBangs(a.id)=1
%\end{lstlisting}

\begin{figure}
	\centering
	\subfloat[Response Time]{
	\includegraphics[width=0.35\textwidth]{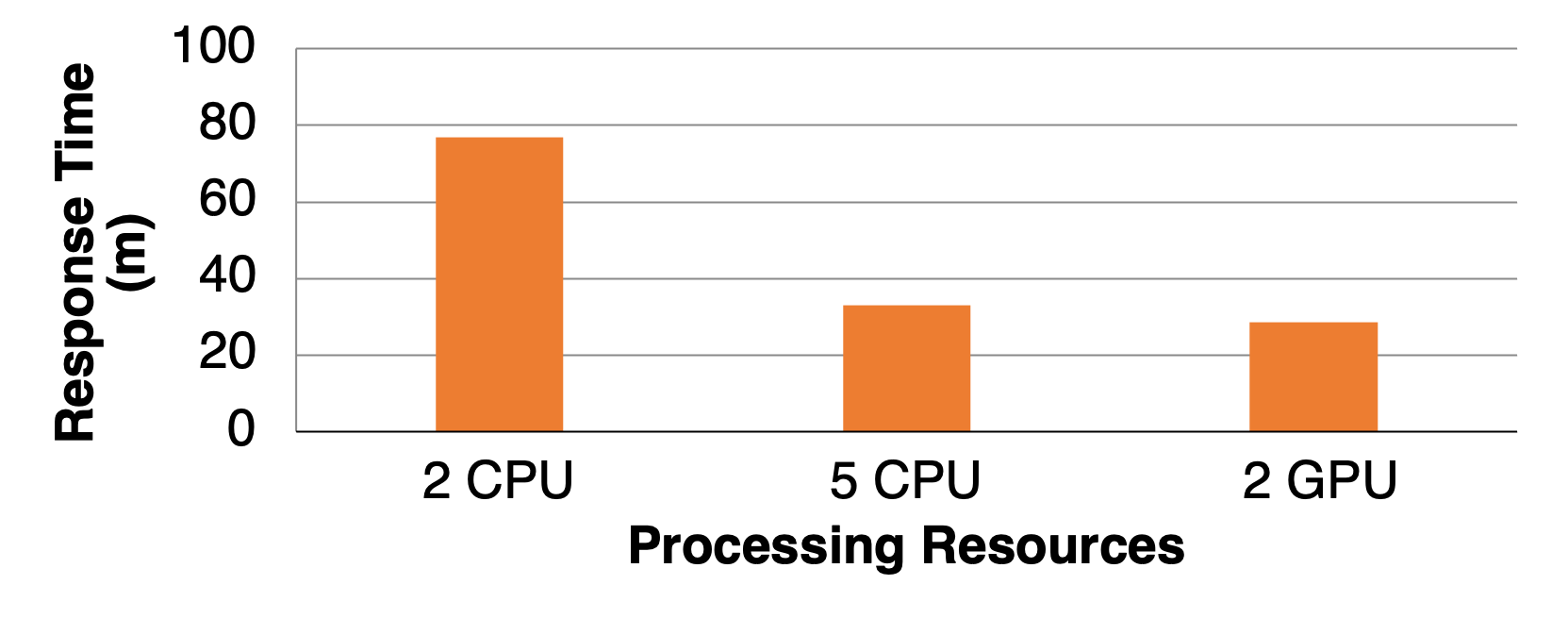}
	\label{query3-results}
}
\subfigurespace
\subfloat[Cost Calculations]{\small
	\label{table:amazonEC2calsq3}
		\begin{tabular}{ccccccc} % <-- Alignments: 1st column left, 2nd middle and 3rd right, with vertical lines in between
			\\
			\centering \textbf{Case} & \textbf{Min} & \textbf{CPU} & \textbf{\$/min} & \textbf{GPU} & \textbf{\$/min} & \textbf{Total}\\
			\hline
			2 CPU & 77 & 2 & \$1.34 & 0 & \$0 & \$1.34 \\
			5 CPU & 34 & 5 & \$1.48 & 0 & \$0 & \$1.48 \\
			2 GPU & 29 & 0 & \$0 & 2 & \$2.96 & \$2.96 \\
		\end{tabular}
}
\caption{Results for Query \#3}
\end{figure}
This query involves UDFs as selection predicates. We have the same two UDFs used for Q1, but in this case, they are used for filtering the table. 
Figure \ref{query3-results} shows the response time results for query \#3.  The disaggregated configuration has the best response time, although  the five node worker configuration is competitive. In terms of cost, however, the disaggregated configuration is the most expensive to run the query.  This appears to be result of the filtering predicates. The first predicate, {\tt hasEyeglasses()}, eliminates a large volume of records, and hence the second predicate {\tt hasBangs()} will get less tuples to evaluate. In this case, the GPU assigned to {\tt hagBangs()} is sitting idle, but still cost money to have it provisioned.  This indicates that there is a need to further consolidate nodes in a disaggregated engine.  

\subsubsection{Query \#4}
%\begin{lstlisting}[language=SQL]
%	select id, smile, isometric,
%	molecular_weight(id) as weight
%	from PubChem
%	where molecular_weight(id) > 437.9
%\end{lstlisting}

\begin{figure}
	\centering
	\subfloat[Response Time]{
	\includegraphics[width=0.35\textwidth]{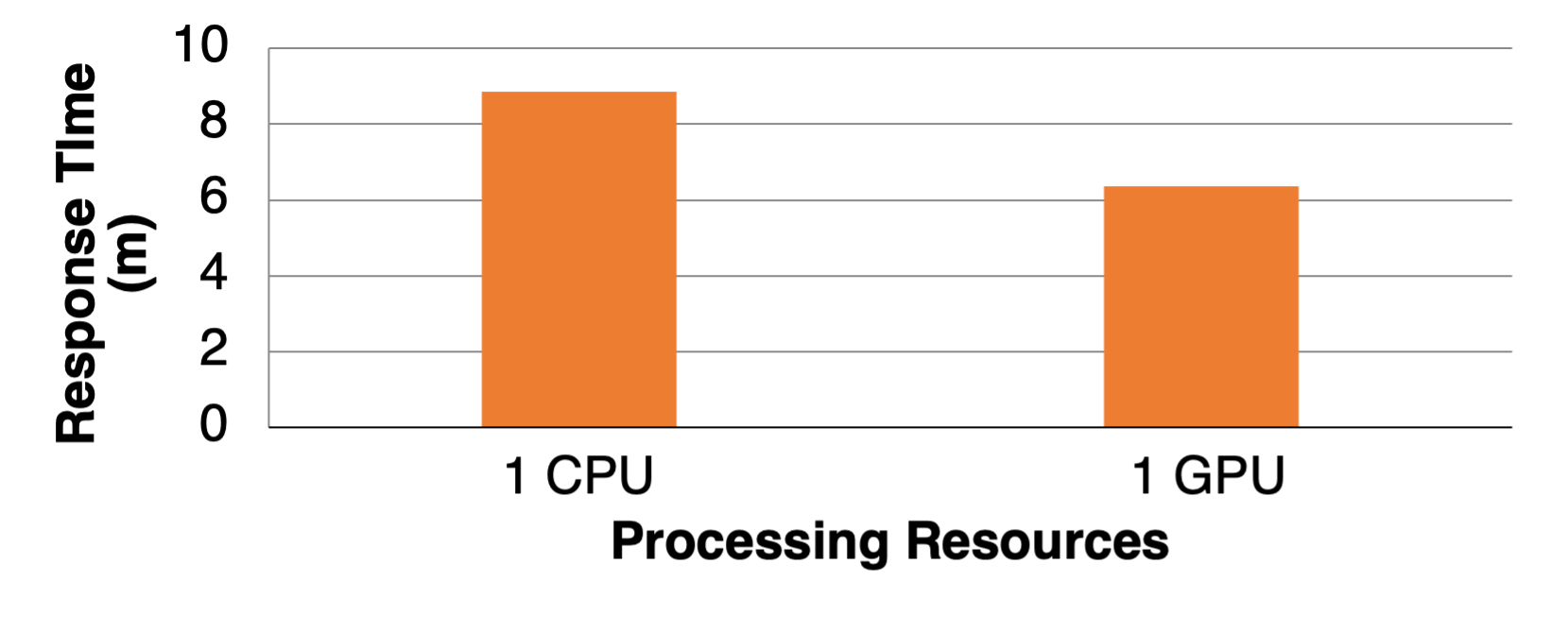}
	\label{query4-results}
}
\subfigurespace
\subfloat[Cost Calculations]{\small
	\label{table:amazonEC2calsq4}
		\begin{tabular}{ccccccc} % <-- Alignments: 1st column left, 2nd middle and 3rd right, with vertical lines in between
			\\
			\centering \textbf{Case} & \textbf{Min} & \textbf{CPU} & \textbf{\$/min} & \textbf{GPU} & \textbf{\$/min} & \textbf{Total}\\
			\hline
			1 CPU & 9 & 1 & \$0.08 & 0 & \$0 & \$0.08 \\
			1 GPU & 7 & 0 & \$0 & 1 & \$0.36 & \$0.36 \\
		\end{tabular}
}
\caption{Results for Query \#4}
\end{figure}
This is a range query that uses the UDF for molecular weight to filter out molecules.  As with case for Q2, the disaggregated configuration has the best response time (Figure \ref{query4-results} but costs more to evaluate (Figure \ref{table:amazonEC2calsq4})). This is due in part to the small size of the data set. 

%Figure \ref{query4-results} shows results for query \#4.  When the  query is run on one high-memory instance, it get the slowest response time at about 9 minutes where as accelerators take about 7 minutes. GPUs reduces time with a speedup of about 1.4x faster. A comparison using monetary cost of the resources shows that the CPU is cheaper by \$0.28 which makes the accelerated intance cost almost 350\% more.

\subsubsection{Query \#5}

%\begin{lstlisting}[language=SQL]
%	select id, smile, isometric,
%	molecular_weight(id) as weight
%	from PubChem
%	where molecular_weight(id) > X and exact_mass(id) > 200
%\end{lstlisting}

\begin{figure}
	\centering
	\subfloat[Response Time]{
	\includegraphics[width=0.35\textwidth]{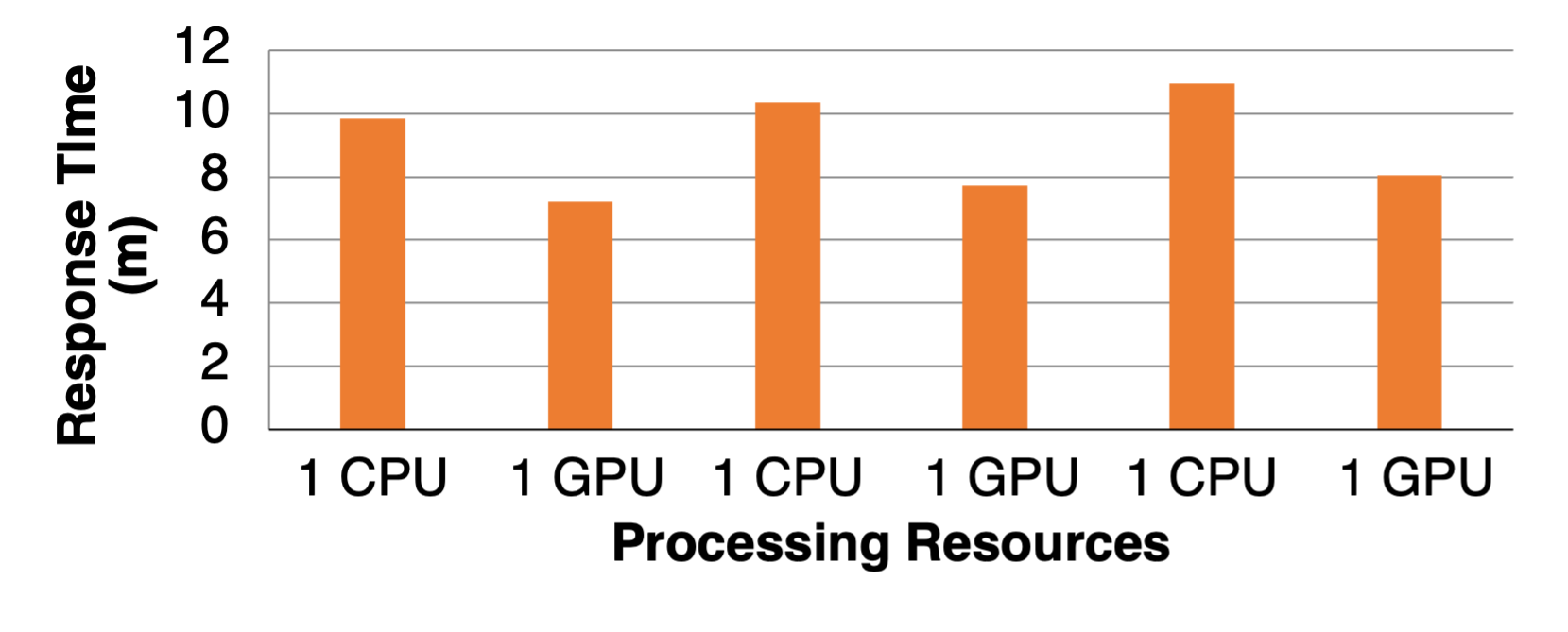}
	\label{query5-results}
}
\subfigurespace
\subfloat[Cost Calculations]{\small
	\label{table:amazonEC2calsq5}
	\resizebox{\columnwidth}{!}{
		\begin{tabular}{ccccccc} % <-- Alignments: 1st column left, 2nd middle and 3rd right, with vertical lines in between
			\\
			\centering \textbf{Case} & \textbf{Min} & \textbf{CPU} & \textbf{\$/min} & \textbf{GPU} & \textbf{\$/min} & \textbf{Total}\\
			\hline
			1 CPU - SP 10\% & 10 & 1 & \$0.09 & 0 & \$0 & \$0.09 \\
			1 GPU - SP 10\%& 8 & 0 & \$0 & 1 & \$0.41 & \$0.41 \\
			\hline
			1 CPU - SP 20\% & 11 & 1 & \$0.10 & 0 & \$0 & \$0.10 \\
			1 GPU - SP 20\%& 8 & 0 & \$0 & 1 & \$0.41 & \$0.41\\
			\hline
			1 CPU - SP 30\% & 11 & 1 & \$0.10& 0 & \$0 & \$0.01 \\
			1 GPU - SP 30\%& 9 & 0 & \$0 & 1 & \$0.46 & \$0.46\\
		\end{tabular}
	}
}
\caption{Results for Query \#5}
\end{figure}
This is another range query, but this time around we varied the selectivity of the  predicate  to filter 10\%, 20\%, and 30\% of the tuples. 
As with  all previous cases for table PubChem, disaggregated operation provides better response time but cost more to run. 

%For query \#5 X was consider using 10\%, 20\% and 30\% which where the values 437.9, 497.6 and 732 respectively.

%The results for the high-memory instance get the slowest response time at about 10 to 11 minutes, where as accelerators take about 8 to 9 minutes. GPUs reduces time with a speedup of about 1.34x to 1.37x faster. A comparison using monetary cost of the resources shows that the CPU is cheaper by \$0.31 to \$0.36 which makes the accelerated intance cost almost 310\% to 360\% more.

\subsection{Join Queries}
\subsubsection{Query \#6}
%\begin{lstlisting}[language=SQL]
%	select a.id, hasEyeglasses(a.id), b.address
%	from celeba as a inner join customer as b
%	on(a.id=b.id)
%	where b.id>20 and hasEyeglasses(a.id);
%\end{lstlisting}

\begin{figure}
	\centering
\subfloat[Response Time]{
	\includegraphics[width=0.35\textwidth]{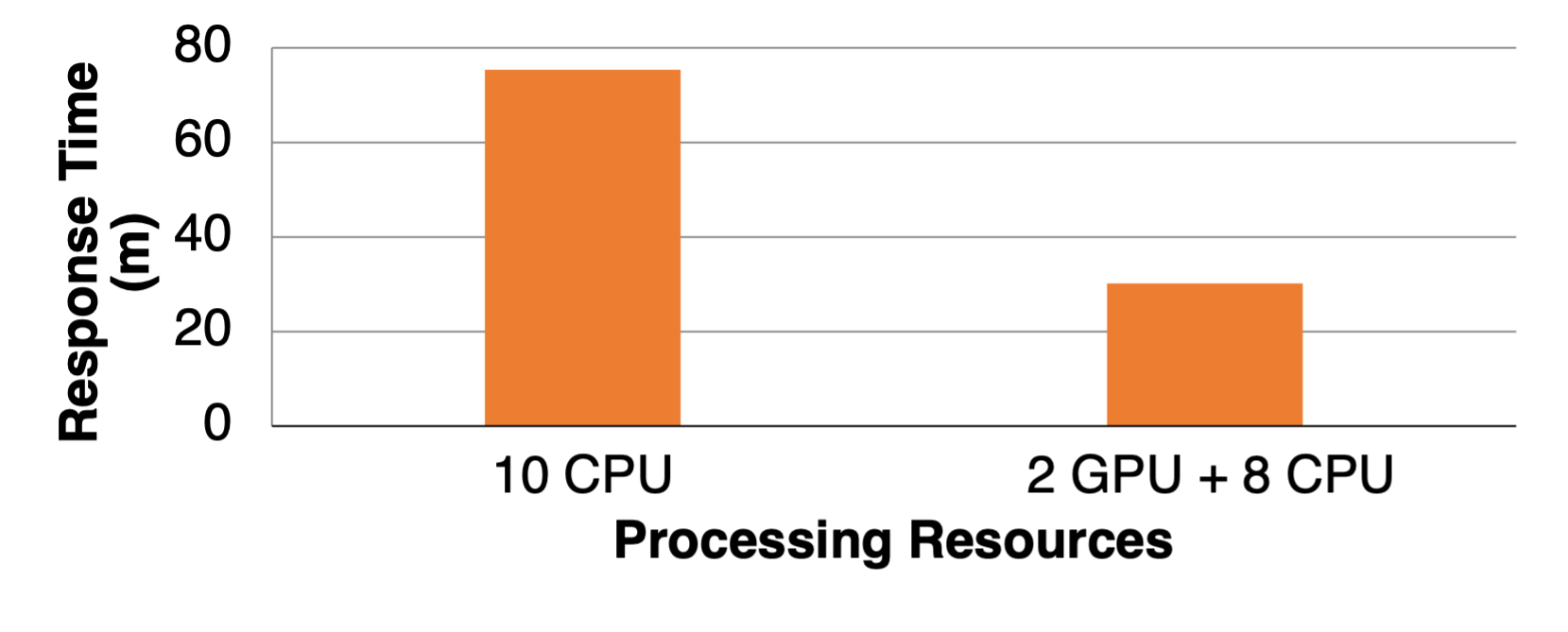}
	\label{query6-results}
}
	\subfigurespace
	\vspace{1mm}
\subfloat[Cost Calculations]{\small
	\resizebox{\columnwidth}{!}{
		\begin{tabular}{ccccccc} % <-- Alignments: 1st column left, 2nd middle and 3rd right, with vertical lines in between
			\\
			\centering \textbf{Case} & \textbf{Min} & \textbf{CPU} & \textbf{\$/min} & \textbf{GPU} & \textbf{\$/min} & \textbf{Total}\\
			\hline
			10 CPU & 76 & 10 & \$6.61 & 0 & \$0 & \$6.61 \\
			2 GPU + 8 CPU & 31 & 8 & \$2.16 & 2 & \$3.16 & \$5.32 \\
		\end{tabular}
	\label{table:amazonEC2calsq6}
}
}
\caption{Results for Query \#6}
\end{figure}
This is a join query featuring a where clause with a UDF predicate. The query combines image data with customer data, after performing the filtering with one of the  predicates
used in Q3. Figure \ref{query6-results} shows results for query \#6.  In this case, a disaggregated configuration with eight (8) high-memory workers, and two (2) GPU workers
performs 2.5x better that  a shared-nothing configuration with ten (10)  high-memory workers. Moreover, this configuration is also cheaper by \$1.29.

%When the  query is run on one high-memory instance, it get the slowest response time at about 76 minutes where as accelerators take about 31 minutes. GPUs reduces time with a speedup of about 2.5x faster. A comparison using monetary cost of the resources shows that using accelerators is cheaper by \$1.29 which  makes it a about 20\% cheaper than using regular instances.

%\subsection{Analysis, Discussion, \& Lessons }
\subsection{Discussion}
From the previous results, we can see that a disaggregated engine, such as ArcaDB, can improve response time for queries by exploiting accelerated instances.
This improvement varied from 1.34x to 3.5x times at its best. The results showed significantly better response times when handling images. This shows that complex formats benefit greatly from accelerated instances since the cost per object to handle them is reduced significantly. 
In contrast, if the data sets are small or if UDF are applied over small objects the performance gains might be marginal. 
%In fact, for alphanumeric datasets it is 
%likely the case that disaggregated systems show little improvement.  
Thus, our preliminary results indicate that these disaggregated engines will excel on workloads that include images, video, and other multimedia files. Also, scientific workloads that need to process big files could also benefit from disaggregation. String-based workloads 
did not show much improvement, although the strings we used are not as long or complex as those used in other domains such as genomics sequencing.  The use of ArcaDB for interactive queries is still an open question. 

As part of our next works, we  need to improve ArcaDB and compare its performance directly with other systems, such as Spark and Hive. 
Also, we still need to study how disaggregated engines perform on user-defined aggregate queries over large data sets, particularly if the data is organized in columnar fashion. 
We also need to study how  using other accelerators (e.g., FPGa), specialized instances (e.g., IO Optimized instances), or other operator implementations may improve response time for other formats and datasets. The hash join operation, for example, benefits greatly indirectly from accelerators when the data to be joined has a complex format but can be filtered before the query (see results for Q6). 
Accelerated instances may participate in previous operators that could reduce the number  items to join, or drop complex attributes.
% so that the plan is expedited.
%First and foremost, performance evaluations occurs by grouping different resources and comparing their results in terms of response time and cost.
%The market regaring data lakes is limited and there are no product currently that support non-structure data query.

%All queries show improvement in response time when accelerated computing instances are used. This improvement varied from 1.34x to 3.5x times at its best. Eventhough accelerated intances made all the queries faster, they showed significantly better response times handleling images. This shows that complex format benefit greatly from accelerated instances since the cost per object to handle it is reduced significatly. In the other hand this also may suggest that using other accelerators, specialized intances or other implementations, may improve response time for other formats and datasets. The join operation ofr example benefics greatly from accelerators when the data involved has a complex format. Accelerated intance may participate in previous operation that could reduce the participating items or their complexity so that the plan is expedited as a whole. Although having less data to hanlde as the query progresses in excution would mean that the query will run in less time, smaller dataset obtain a marginal benefit when using accelerators. In this case the client must determine if the substancial improvement is worth it.
%
%This shows that certain operation, in this case string management, might obtain better results if other libraries, languages or intances are used instead.

Notice that using  ``on demand'' accelerated instances can reduce query time, but might be more expensive. Thus, we are trading monetary cost for  having queries run faster. But this monetary cost is only relevant when host services are rented on demand from  a cloud. Theses costs might be dismissed if ArcaDB is used on premise, where the capital expenses have already been incurred. In such case, taking advantage of the available resources and reducing response time is a better criteria to evaluate the performance of a disaggregated system like ArcaDB. In a market where fungible cloud resources yield similar results, we need to explore different resource allocations to identify which combination is better suited for each query workload. In fact, we might need to explore multi-objective optimization to scope resource usages. For example, we might have a query optimizer that receives as monetary budget and minimizes response time subject to the budget. In general, multiple objectives like costs, budgets and resources could be considered for query optimization. New heuristics and policies  may be implemented to develop  plans and resource allocations that satisfy all constrains.

%% file: relatedwork.tex
\section{Related Works\label{relworks}}
Previous efforts have pioneered the use of accelerators for query processing \cite{10.14778/3551793.3551809, 10.14778/3484224.3484229, 10.1145/3318464.3380595}. The works in \cite{10.1145/1376616.1376670, 10.1145/2236584.2236592, 10.14778/3436905.3436927}  explored the use of a GPU to run join operators, obtaining significant performance improvements over CPU-based joins. GPU-based acceleration has also being studied for aggregation  \cite{Karnagel2015OptimizingGG}. The work in \cite{10.1145/3514221.3517842} explored the use of GPU for sorting operations, obtaining significant performance gains. FPGAs also been explored as accelerators for database workloads. Several efforts explored the use of FPGA to accelerate string matching operators \cite{10.1145/3035918.3035954}, group by queries \cite{10.1145/2933349.2933360}, columnar scans \cite{DBLP:conf/vldb/LisaUHLN018}, and sorting \cite{10.1145/3399666.3399897}. ArcaDB differs 
from these systems in two main ways. First, each solution is designed to work (or is tested) on a single node system. The operator off-loading  happen between the CPU and accelerator (GPU, TPU, or FPGA) in this single machine. Second, the solutions are homogeneous in terms of software stack and monolithic in terms of query engine. ArcaDB, on the hand, works on a cluster of commodity machine that include accelerators, general purpose nodes, high-memory modes, and other special-purpose machines. Moreover, ArcaDB supports heterogeneous UDFs and UDTs than can be implemented in various languages to support each query.

There is an extensive amount of work in query processing in areas such as client-server systems \cite{mf:trade}, integration systems \cite{DBLP:conf/sigmod/RodriguezR00}, and data analytics  \cite{MapReduce, Thusoo:2009:HWS:1687553.1687609, 10.1145/2723372.2742797, 8731547}. In these efforts, the bulk of the work has been performed to support distributed join processing \cite{10.1145/3325135}, transaction processing \cite{10.14778/3415478.3415560}, and data caching \cite{10.1145/2670979.2670985, Li:EECS-2018-29}. These efforts differ from ArcaDB mainly because these engines are monolithic in nature, or place the same query plan everywhere for workers to evaluate the plan in their assigned data shards. However, in the presences of accelerators, it makes sense to disagregate the engine, distributing query operators  in an asymmetric fashion to nodes that better fit their performance requirements. ArcaDB excels in this aspect, enabling individual allocation of resources on a per-operator basis. Our results show the advantages on such an approach. 

Recently, there has been a lot of interest on serverless computing. However, these solutions have been criticized  for their lack of inter-function communication and overheads to run a pipeline of operations \cite{jonas2019cloud}. ArcaDB itself is not a serverless system, but it can be integrated with such systems. Functions can be executed as a UDF over a virtual table that provides the data.  ArcaDB can help drive the serveless system, providing data processing, caching,  or  chaining function calls as successive query calls.  

%% file: conclusions.tex
\section{Conclusions\label{conclusions}}
In this paper, we presented ArcaDB, a disaggregated query engine that applies container technology to place operators and UDFs at compute nodes that fit their performance needs. This novel architecture, employs the use of heterogeneous resources to support query execution. Each operator/UDF runs on a preferred resource. These operators allocations are performed heuristically by the coordinator. ArcaDB ensures that the operator gets picked up by the appropriate workers. With this arrangement, we can maximize the use of special-purpose instances, reduce query response time, and keep the per-query costs under control.  We implemented a prototype version of ArcaDB using Java, Python, Docker containers, and other supporting open-source tools. We also completed a preliminary performance study of this prototype, using image and scientific data sets. This study shows that ArcaDB can speedup query performance up to a factor of 3.5x in comparison with a shared-nothing, symmetric arrangement. Thus, ArcaDB can help users better meet the performance requirements of their applications by utilizing heterogenous resources wherever they need.

%% file: acknowledgment.tex
\section{Acknowledgments}
%This is the product from countless hours of work and reunions. It worked. 
This material is based upon work supported by the National Science Foundation under Grant OIA-1849243. Any opinions, findings, and conclusions or recommendations expressed in this material are those.
Research reported in this publication was also supported by the National Library of Medicine, of the National Institutes of Health under award number R15LM012275. The content is solely the responsibility of the authors and does not necessarily represent the official views of the National Institutes of Health. 